\documentstyle[11pt]{article}
\begin{document}

\title{Universal Similarity Factorization Equalities over  Real Clifford Algebras }

\author{Yongge  Tian \\
Department of Mathematics and Statistics \\
Queen's University \\
Kingston,  Ontario,  Canada K7L 3N6\\
{\tt e-mail:ytian@mast.queensu.ca}}
\date{}

\date{}

\maketitle

\noindent {\bf Abstract.} \ {\small A variety of universal similarity
factorization equalities over real Clifford algebras ${\cal R}_{p,q}$ are
established. On the basis of these equalities, real, complex and
quaternion matrix representations  of elements in ${\cal R}_{p,q}$
can  be explicitly determined. \\

}  \\

\noindent {\bf 1. \ INTRODUCTION } \\

The aim of this paper is to establish  universal similarity factorization
equalities between elements of  real Clifford algebras  $ {\cal R}_{p,q} $
and matrices with elements in $ {\cal R}, \ {\cal C}$ and $ {\cal H},$ where
$ {\cal R}, \ {\cal C} $ and  $ {\cal H}$  stand for  real number field,
  complex  number field and quaternion skew field, respectively. A direct
  motivation for considering this problem comes from the following basic
  universal similarity factorization equality for complex numbers:
$$
\left[ \begin{array}{cr} 1  & i \\ -i &  -1  \end{array} \right]\left[ \begin{array}{cc}  a + bi  & 0 \\ 0 & a - bi \end{array} \right]
\left[ \begin{array}{cr} 1  & i \\ -i &  -1  \end{array} \right]=  \left[ \begin{array}{cr}  a  &  -b \\ b & a  \end{array} \right]. \eqno (1.1) 
$$  
This equality clearly reveals three fundamental facts on the field of
complex numbers

(a) \ ${\cal C}$  is algebraically isomorphic to the matrix algebra ${\cal A} = \left\{ \left. \left[ \begin{array}{cr}  a  &  -b  \\ b & 
a  \end{array} \right]  \ \right| \ a, \ b \in {\cal R}  \ \right\}$ through the bijective  map $ \phi: a + bi  \longrightarrow \left[ \begin{array}{cr}  a  &  -b \\ b & a  \end{array} \right]$. 
 
(b) \ Every complex number $ p = a + bi $ has a faithful matrix representation $ \phi(p) = \left[ \begin{array}{cr} a & -b \\ b & a  \end{array} \right]$ over the real number field $ {\cal R}$.

(c) \ All real matrices of the form $\left[ \begin{array}{cr} a & -b \\ b & a  \end{array} \right]$ can uniformly be diagonalized over  the complex number field $ {\cal C}$. 

\medskip

Following Eq.(1.1), a natural equation can directly be asked: Can we extend Eq.(1.1) to any real Clifford algebra ${\cal R}_{p,q}$? The answer to this question is positive. In this paper, we shall present a set of general methods for establishing such kinds of  universal similarity factorization equalities over all real Clifford algebras ${\cal R}_{p,q}$. 

\medskip

 As is well known, the real Clifford algebra $ {\cal R}_{p,q} $ is an
 associative algebra, with identity 1, defined on  $p+q =n$  generators
 $ e_1, \ e_2, \  \cdots,  \ e_n $ subject to   the multiplication laws
 $$
  e_i^2 = \left\{  \begin{array} {l}   +1  \qquad for \ i = 1, \ 2, \ \cdots, \ p,  \\ -1 
 \qquad  for \ i = p+1, \ p+2, \ \cdots, \ p +q = n, \end{array} \right.     \eqno (1.2) 
$$ 
$$ 
 e_ie_j + e_je_i = 0 \qquad  for \ \  i \neq j, \qquad  i,  \ j
 = 1,  \ 2, \ \cdots, \ n.   \eqno (1.3)
$$ 
and $ e_1e_2 \cdots e_n  \neq \pm1$. In that case $ {\cal R}_{p,q} $ is
spanned as a $ 2^n$-dimensional vector space with a basis $ \{ e_A\}$,
where the multiindex $ A $ ranges all naturally ordered subsets of the first
positive integer set $ \{ 1, \ 2, \ \cdots, \ n \}$; the basis
element $ e_A $, where $ A = ( \, i_1, \ i_2, \ \cdots,  \ i_k \,  ) $ with $ 1 \leq i_1 < i_2 < \cdots < i_k \leq  n$, is defined as the product 
$$
  e_A = e_{(i_1, i_2, \cdots, i_k)} \equiv e_{i_1}e_{i_2} \cdots e_{i_k}.   
$$ 
In particular, $ e_A = 1,$ when $ A = \emptyset$. For simplicity, a brief
notation $  e_{[n]} = e_1e_2 \cdots e_n$ will be adopted in the sequel. The
square of $e_{[n]}$ is 
$$
 e_{[n]}^2 = (-1)^{\frac{1}{2}n(n-1)}e_1^2e_2^2\cdots e_n^2.
$$
Any element $ a \in {\cal R}_{p,q}$ can be expressed as 
$$ 
a = \sum_A a_Ae_A,   \qquad  a_A \in {\cal R},
 $$   
where $A$ ranges all naturally ordered subsets of
$ \{ 1, \ 2, \ \cdots, \ n \}$. We shall also adopt the  following  notation
 in the sequel
$$
 {\cal R}_{p,q} = { \cal R }\{ \,  e_1, \ \cdots, \ e_p, \  \varepsilon_1,
 \ \cdots, \ \varepsilon_q \ | \ e_i^2 = 1, \ \ \varepsilon_j^2 =-1, \ \
 i = 1, \ \cdots, \ p, \ \  j = 1, \ \cdots, \ q \, \},
 $$ 
or simply
$$
{\cal R}_{p,q} = { \cal R }\{ \,  e_1, \ \cdots, \ e_p, \  \varepsilon_1, \
 \cdots, \ \varepsilon_q \, \},
 $$
and use
$$
{\cal R}_{p,q}^{m \times n} = { \cal R }^{m \times n} \{ \,  e_1, \ \cdots, \ e_p, \  
\varepsilon_1, \ \cdots, \ \varepsilon_q  \, \}
$$ 
to stand for  the collection of all $m\times n$ matrices over
${\cal R}_{p,q}.$ 

\medskip

As to  the algebraic structure of the Clifford algebra $ {\cal R}_{p,q}$
with $ p + q = n $, it is well-known(see, e.g., \cite{1},  \cite{8},  \cite{9},  \cite{11})
that $ {\cal R}_{p,q}$ satisfies the following algebraic isomorphisms
$$ 
 {\cal R}_{p,q} 
\simeq 
\left\{ 
 \begin{array} {l} {\cal R}(2^{n/2}) \ \ \ \ \ \ \ \ \ if \ q-p \equiv 0, \ 6 \pmod{8}, \\
{\cal C}(2^{(n-1)/2})  \ \ \ \ \   if \ q-p \equiv 1, \ 5 \pmod{8}, \\
{\cal H}(2^{(n-2)/2})  \ \ \ \  if \ q-p \equiv 2, \ 4 \pmod{8}, \\
^2{\cal H}(2^{(n-3)/2})  \ \ \  if \ q-p \equiv 3  \ \ \ \pmod{8}, \\
^2{\cal R}(2^{(n-3)/2}) \ \ \  if  \ q-p \equiv 7  \ \ \ \pmod{8}, 
 \end{array} \right.  \eqno (1.4)
 $$ 
where $ {\cal R}(s), \  {\cal C}(s), \  {\cal H}(s) $ stand for the matrix algebras $ {\cal R}^{s \times s}, \  {\cal C}^{s \times s}, \  {\cal H}^{s \times s}$, respectively, and  $ ^2{\cal R}(s), \ ^2{\cal H}(s)$  stand for the matrix algebras
$$
  ^2{\cal R}(s) = \left\{ \left. \left[ \begin{array}{cc}  A  & O   \\ O & B \end{array} \right] \
  \right|  \ A, \ B \in {\cal R}^{s \times s}  \right\}, \ \ \  ^2{\cal H}(s) = \left\{  \left. \left[ \begin{array}{cc}  A & O   \\ O & B \end{array} \right] \  \right|  \ A, \ B \in {\cal H}^{s \times s}  \right\}.
$$ 

For some low values of $ p $ and $ q $, Eq.(1.4) can be expressed as 
$$
 {\cal R}_{1,0}  \ \simeq  \ ^2{\cal R}, \qquad {\cal R}_{0,1}  =  {\cal C},  \eqno (1.5) 
$$
$$
 {\cal R}_{2,0}  \ \simeq  \  {\cal R}^{2 \times 2}, \qquad {\cal R}_{1,1} \  \simeq  \ {\cal R}^{2 \times 2}, \qquad {\cal R}_{0,2}  = {\cal H},     \eqno (1.6) 
$$
$$ 
 {\cal R}_{3,0}  \ \simeq  \ {\cal C}^{2 \times 2}, \qquad {\cal R}_{2,1}  \ \simeq  \ ^2{\cal R}^{2 \times 2},
\qquad {\cal R}_{1,2} \  \simeq   \  {\cal C}^{2 \times 2},  \qquad {\cal R}_{0,3} \  \simeq  \ ^2{\cal H},  
 \eqno (1.7)
 $$
$$ 
 {\cal R}_{4,0} \  \simeq  \ {\cal H}^{2 \times 2}, \ \ \  {\cal R}_{3,1} \  \simeq  \ {\cal R}^{4 \times 4},
\ \ \ {\cal R}_{2,2} \  \simeq  \  {\cal C}^{4 \times 4}, \ \ \  {\cal R}_{1,3} \  \simeq  \ {\cal H}^{2 \times 2}, \ \ \ {\cal R}_{0,4} \  \simeq \ {\cal H}^{2 \times 2}.   \eqno (1.8)
$$
In addition, according to the periodicity theorem on Clifford
algebras( see, e.g., \cite{3} and \cite{11}), it is also well-known that
$$
 {\cal R}_{p+8,q}  \ \simeq \   {\cal R}_{p,q}^{16 \times 16}, \qquad {\cal R}_{p,q+8} \  \simeq
  \ {\cal R}_{p,q}^{16 \times 16}  \eqno (1.9) 
$$ 
hold  for all  finite $ p $ and $ q $. 

\medskip

The isomorphisms listed above imply that there exists a one-to-one
correspondence, preserving algebraic operations, between elements of
${\cal R}_{p,q} $ and matrices with elements in $ {\cal R}$ or  ${\cal C}$
or ${\cal H}$, what we shall do in the present paper is to explicitly
establish universal similarity factorization equalities between elements of
 ${\cal R}_{p,q}$ and matrices with elements in ${\cal R},$ ${\cal C}$, and
 ${\cal H}.$ 
\medskip

A key tool used in the sequel is given below. 
\medskip

\noindent {\bf Lemma 1.1.} \ {\em Let $ {\cal A} $ be an algebra over
 an arbitrary field $ {\cal F},$ and let $ M_n({\cal A} )$ be the
 $ n \times n$ total matrix algebra with elements in $ {\cal A},$ and with
 its basis $\{ \tau_{ij} \} $ satisfying the multiplication rules
 $$
  \tau_{ij} \tau_{st} = \left\{ \begin{array}{c} \tau_{it}  \qquad j = s,  \\ 0 \qquad j \neq s, \end{array} \right. \qquad  for \ \  i, \ j, \ s, \ t = 1, \ 2, \ \cdots, \  n. \eqno (1.10) 
$$   
Then any $ a = \sum^n_{i,j = 1}  a_{ij}\tau_{ij} \in  M_n({\cal A}),$ where $ a_{ij} \in {\cal A},$  satisfies the following universal similarity  factorization equality
$$
 P  \left[ \begin{array}{cccc} a &  &  &  \\ & a & & \\  &  & \ddots  & \\
 &  &  & a  \end{array} \right] P^{-1} = \left[ \begin{array}{cccc} a_{11} & a_{12} & \cdots & a_{1n} 
 \\ a_{21} & a_{22} & \cdots & a_{2n} 
 \\ \cdots &  \cdots & \cdots &  \cdots \\
 a_{n1} & a_{n2} & \cdots & a_{nn}  \end{array} \right],  \eqno (1.11)
$$  
where $ P $ has the following independent form 
$$ P =  P^{-1} = \left[ \begin{array}{cccc} \tau_{11} & \tau_{21} & \cdots & \tau_{n1} \\ 
 \tau_{12} & \tau_{22} & \cdots & \tau_{n2} \\
\cdots &  \cdots & \cdots  &  \cdots \\
 \tau_{1n} & \tau_{2n} & \cdots & \tau_{nn}   \end{array} \right].
  \eqno (1.12)
$$ } 

The correctness of this result can directly be verified by multiplying the
 left-hand side of Eq.(1.11). The significance of this result is in that if an algebra $ {\cal M} $ is known to be algebraically isomorphic to an $ n \times n $ 
total matrix algebra over $ {\cal A} $, then there exists an independent invertible matrix $ P $  over 
$ {\cal M} $ such that  $ P( aI_n )P^{-1} \in {\cal A }^{n \times n}$ holds for all $ a \in {\cal M}$. 
The most part of  the results in the paper are established 
through this basic identity. \\

\noindent {\bf 2.  UNIVERSAL SIMILARITY EQUALITIES OVER $ {\cal R}_{p,q } $ WITH $ p +q \leq 8$ } \\

This section is divided into eight subsections. In the first four of which, we present the universal similarity factorization equalities over $ {\cal R}_{p,q }$ corresponding to $ p +q \leq 4$ with proofs, and then list the  universal  similarity factorization equalities over $ {\cal R}_{p,q }$ corresponding to $ 5 < p +q \leq 8$.\\

\noindent {\bf 2.1. The Cases ${\cal R}_{p,q}$  with $ p +q = 1$ } \\

The two algebras $ {\cal R}_{p,q }$ corresponding to  $ p +q = 1$ are ${\cal R}_{1,0 }$, the hyperbolic numbers, and ${\cal R}_{0,1},$ the complex numbers, respectively. The two fundamental universal similarity factorization  equalities over them are very simple but crucial for  the subsequent results.

\medskip

\noindent {\bf Theorem 2.1.1.} \ {\em Let $ a= a_0 + a_1e_1 \in {\cal R}_{1,0}$ be given, where $ a_0, \  a_1 \in {\cal R}, \ 
e^2_1 = 1,$ and denote its conjugate $ \overline{a} = a_0 - a_1 e_1 $. Then $ a $ and $ \overline{a} $ satisfy the following universal similarity factorization equality 
$$
 P_{1,0} \left[ \begin{array}{rr}  a  & 0   \\ 0 &  \overline{a} \end{array} \right] P_{1,0}^{-1}
= \left[ \begin{array}{cc}  a_0 + a_1 & 0  \\ 0   & a_0  - a_1  \end{array} \right] \equiv \phi_{1,0}(a)
 \in \, ^2{\cal R },   \eqno (2.1.1) 
$$  
where $P_{1,0}$  and $ P_{1,0}^{-1}$  have the independent forms $($no relation with $ a )$ 
$$ 
P_{1,0} = \frac{1}{2} \left[ \begin{array}{cc} 1 + e_1  & -(1 - e_1)  \\  1 - e_1 &  1+ e_1  \end{array} \right],  \qquad  P_{1,0}^{-1}  = \frac{1}{2} \left[ \begin{array}{cc} 1+e_1  & 1 - e_1 \\ -(1 - e_1)  &  1+ e_1  \end{array} \right]. \eqno (2.1.2) 
$$ }
 \noindent {\bf Proof.} \ Let $ s = 1+e_1$. Then $ \overline{s} =  1 - e_1$, and both of them satisfy $ s^2 = 2s, \ \overline{s}^2 = 2\overline{s}$ and $ s \overline{s} =  \overline{s}s = 0.$  Thus it follows that 
\begin{eqnarray*}
 P_{1,0} \left[ \begin{array}{cc}  a  & 0   \\ 0 &  \overline{a} \end{array} \right] P_{1,0}^{-1}
& = & \frac{1}{4} \left[ \begin{array}{cr} s  & -\overline{s}  \\  \overline{s}  & s \end{array} \right]
 \left[ \begin{array}{cc}  a  & 0 \\ 0 &  \overline{a} \end{array} \right] \left[ \begin{array}{cc} s  & \overline{s}  \\  -\overline{s}  & s \end{array} \right] \\
& = &  \frac{1}{4} \left[ \begin{array}{cc} sas + \overline{s} \, \overline{a} \, \overline{s} &
 sa \overline{s} - \overline{s} \, \overline{a}s \\  \overline{s}as &  \overline{s} a \overline{s} + s\overline{a}s    \end{array} \right] \\
 & = &  \frac{1}{4} \left[ \begin{array}{cc} as^2 + \overline{a} \,\overline{s}^2   & 0 \\ 0  &
 a\overline{s}^2 + \overline{a}s^2 \end{array} \right] =  \frac{1}{2} \left[ \begin{array}{cc} as + \overline{a} \, \overline{s}   & 0 \\ 0  & a\overline{s} + \overline{a}s  \end{array} \right],
\end{eqnarray*}
where the two  nonzero terms in it are 
$$
 as + \overline{a} \, \overline{s} = (a_0 + a_1e_1)( 1 + e_1) + (a_0 - a_1e_1)( 1 - e_1) = 2( a_0 + a_1),
$$ 
$$
a\overline{s} + \overline{a}s = (a_0 + a_1e_1)( 1 - e_1) + (a_0 - a_1e_1)( 1 + e_1) = 2( a_0 - a_1).
$$
Hence we have Eq.(2.1.1).   \qquad $\Box $ 

\medskip

It is easily seen that 
$$ 
  \overline{\overline{a}} = a,   \qquad\
  \overline{ a + b } = \overline{ a } + \overline{ b }, \qquad
\overline{ab}=\overline{a}\overline{b}, \qquad \overline {\lambda a}
 = \overline { a \lambda } = \lambda \overline{a}
 $$  
hold for all $ a, \ b \in {\cal R}_{1,0}, \ \lambda \in {\cal R}$. Thus by Eq.(2.1.1) it follows that 

{\rm (a)} \ $ a=b \ \Longleftrightarrow  \ \phi_{1,0}(a) = \phi_{1,0}(b).$ 

{\rm (b)}  \ $ \phi_{1,0}(a + b ) =  \phi_{1,0}(a) + \phi_{1,0}(b), \qquad
 \phi_{1,0}(ab)= \phi_{1,0}(a) \phi_{1,0}(b), \qquad
 \phi_{1,0}( \lambda a)= \lambda \phi_{1,0}(a).$

{\rm (c)} \ $ \phi_{1,0}(1 )= I_2. $ 

{\rm (d)} \ $ a = \frac{1}{4}[ \, 1 + e_1, \ 1 - e_1 \,]
\phi_{1,0}(a)[ \, 1 + e_1, \ 1 - e_1 \, ]^T .$

{\rm (e)} \ $ {\rm det}[ \phi_{1,0}(a) ]=  a_0^2 - a_1^2$ \ \ for all
 $a = a_0 + a_1e_1 \in {\cal R}_{1,0}.$

{\rm (f)} \ $ a $ is invertible  $ \Longleftrightarrow \phi_{1,0}(a) $ is
invertible, in which case,  $ \phi_{1,0}(a^{-1}) = \phi_{1,0}^{-1}(a).$ 

{\rm (g)} \ $ p_a(a ) = 0,$ where $ p_a( \lambda )
= {\rm det}[ \, \lambda I_2 - \phi_{1,0}(a) \,].$  \\
These properties imply that through the bijective map 
$$
 \phi_{1,0}: \ a \in{\cal R}_{1,0} \longrightarrow \phi_{1,0}(a) \in \, ^2{\cal R}, 
 $$ 
the Clifford algebra  ${\cal R}_{1,0} $ is algebraically isomorphic to the matrix algebra $ ^2{\cal R}$, and $ \phi_{1,0}(a)$ is a matrix representation of $ a \in {\cal R}_{1,0}$ in $ ^2{\cal R}$. \\

As to the algebra ${\cal R}_{0,1} = {\cal C }$, the  following result is quite easy to verify . \\
 
\noindent {\bf Theorem 2.1.2.} \ { \em Let $ a= a_0 + a_1\varepsilon_1 \in {\cal R}_{0,1} = {\cal C},$ where $ a_0, \ a_1 \in {\cal R}, \ \varepsilon_1^2 = -1,$ and denote $ \overline{a} = a_0 - a_1\varepsilon_1.$  Then $ a $ and $ \overline{a} $ satisfy the following universal similarity factorization equality 
$$
 P_{0,1} \left[ \begin{array}{cc}  a  & 0   \\ 0 &  \overline{a} \end{array} \right] P_{0,1}^{-1}
= \left[ \begin{array}{cr}  a_0  &  -a_1 \\ a_1 & a_0  \end{array} \right] \equiv \phi_{0,1}(a)
 \in {\cal R }^{ 2 \times 2} ,   \eqno (2.1.3) 
$$  
where $P_{0,1}$  has the independent form  
$$ 
P_{0,1} = P_{0,1}^{-1} = \frac{1}{\sqrt {2}} \left[ \begin{array}{cr} 1  & \varepsilon_1 \\ - \varepsilon_1 &  -1  \end{array} \right].    \eqno (2.1.4)
 $$
Through the bijective map $ \phi_{0,1}: a \longrightarrow \phi_{0,1}(a),$
the algebra $ {\cal R}_{0,1}$ is algebraically isomorphic to the matrix
algebra ${\cal A} = \left\{ \left. \left[ \begin{array}{cr}  a_0  &  -a_1
\\ a_1 & a_0  \end{array} \right]  \ \right| \ a_0, \ a_1 \in {\cal R}  \
\right\}. $ } 

The two equalities in Eqs.(2.1.1) and (2.1.4) can be extended to all matrices over ${\cal R}_{1,0}$ and $ {\cal R}_{0,1}$ as follows.  

\medskip

\noindent {\bf Theorem 2.1.3.} \ {\em Let $ A = A_0 + A_1e_1 \in {\cal R}_{1,0}^{m \times n }, $ where  $ A_0, \ A_1 \in {\cal R}^{m \times n }$  and $ e^2_1 =1.$  Then $ A $ and its conjugate $ \overline{A} = A_0 - A_1e_1 $ satisfy the
 following  universal factorization equality 
$$
 Q_{2m} \left[ \begin{array}{cc}  A  & O   \\ O &  \overline{A} \end{array} \right] Q_{2n}^{-1}
= \left[ \begin{array}{cc}  A_0 +A_1 & O  \\ O & A_0  - A_1  \end{array} \right]  \equiv \Phi_{1,0}(A)
 \in  \,  ^2{\cal R }^{m \times n},  \eqno (2.1.5) 
$$  
where  
$$ 
Q_{2m} = \frac{1}{2} \left[ \begin{array}{cc} (1+e_1)I_m  & -(1 - e_1)I_m  \\  (1 - e_1)I_m &  (1+ e_1)I_m  \end{array} \right],  \qquad  Q_{2n}^{-1}  = \frac{1}{2} \left[ \begin{array}{cc} (1+e_1)I_n  & (1 - e_1)I_n \\ -(1 - e_1)I_n  &  (1+ e_1)I_n  \end{array} \right]. $$ 
In particular, when $ m = n $,  Eq.{\rm (2.1.5)} becomes a universal similarity factorization equality over
 $ {\cal R}_{1,0}$.} 

\medskip

\noindent {\bf Theorem 2.1.4.} \ {\em Let $ A = A_0 + A_1\varepsilon_1 \in {\cal R}_{0,1}^{m \times n} = {\cal C}^{m \times n},$ where $ A_0, \ A_1 \in {\cal R}^{m \times n}, \ \varepsilon_1^2 = -1.$  Then $ A $ and its conjugate  $ \overline{A} = A_0 - A_1\varepsilon_1$  satisfy the following  universal factorization  equality 
$$
 K_{2m} \left[ \begin{array}{cc}  A  & 0   \\ 0 &  \overline{A} \end{array} \right] K_{2n}^{-1}
= \left[ \begin{array}{cr}  A_0  &  -A_1 \\ A_1 & A_0  \end{array} \right] \equiv \Phi_{0,1}( A )  \in {\cal R }^{ 2m \times 2n} ,   \eqno (2.1.6) 
$$  
where 
$$ 
K_{2t} =K_{2t}^{-1} = \frac{1}{\sqrt {2}} \left[ \begin{array}{cr} I_t  & \varepsilon_1I_t \\
 - \varepsilon_1I_t &  -I_t  \end{array} \right], \qquad t = m, \ n.
$$
In particular, when $ m = n $, Eq.{\rm (2.1.6)} becomes a universal similarity factorization equality over
 $ {\cal R}_{0,1} = {\cal C}.$} \\

\noindent {\bf 2.2. The Cases for ${\cal R}_{p,q}$ with $ p + q = 2$} \\

The three algebraic isomorphisms for $ \in {\cal R}_{p,q} $ with $ p + q = 2$ are shown in Eq.(1.6). Based on
Lemma 1.1 and Theorem 2.1.1, we can establish universal similarity
factorization equalities over them as follows. \\

\noindent {\bf Theorem 2.2.1.} \ {\em Let $ a \in {\cal R}_{2,0}
= {\cal R} \{ e_1, \ e_2 \ | \ e_1^2=1, \ e_2^2 =1 \}.$ Then $a$ can factor
as $ a=  a_0 + a_1e_1 + a_2e_2 + a_3e_{12},$ where $a_0$---$a_3 \in {\cal R}
 ,$ and $aI_2$  satisfies the following universal similarity
factorization equality 
$$
 P_{2,0} \left[ \begin{array}{cc}  a  & 0   \\ 0 & a  \end{array} \right] P_{2,0}^{-1}
= \left[ \begin{array}{cc}  a_0 + a_1 &  a_2 + a_3  \\   a_2 - a_3 & a_0  - a_1  \end{array} \right] \equiv \phi_{2,0}(a) \in {\cal R}^{2 \times 2 },   \eqno (2.2.1)
 $$
where $P_{2,0}$  has the independent form  
$$ 
P_{2,0} = P_{2,0}^{-1} = \frac{1}{2} \left[ \begin{array}{cc} 1+e_1  & e_2 - e_{12} \\  e_2 + e_{12} & 
 1 - e_1 \end{array} \right].  \eqno (2.2.2)
 $$ } 
 
\noindent {\bf Proof.} \ By Lemma 1.1, we take the change of basis of ${\cal R}_{2,0}$ as follows 
$$ 
\tau_{11}= \frac{1}{2}( 1+e_1), \ \ \ \tau_{12}= \frac{1}{2}( e_2 + e_{12}), \ \ \  
\tau_{21}= \frac{1}{2}( e_2 - e_{12}), \ \ \ \tau_{22}= \frac{1}{2}( 1 - e_1).  \eqno (2.2.3) 
$$
Then it is not difficult to verify that this new basis satisfies the multiplication laws in Eq.(1.10). In that 
case,  every $ a = a_0 + a_1e_1 + a_2e_2 + a_3e_{12} \in {\cal R}_{2,0}$ can be rewritten as 
$$
a= ( a_0 + a_1)\tau_{11} + ( a_2 + a_3)\tau_{12} + ( a_2 - a_3)\tau_{21} + ( a_0 - a_1)\tau_{22}.
 \eqno (2.2.4)
$$
Substituting Eqs.(2.2.3) and (2.2.4) into Eqs.(1.11) and (1.12),  we directly obtain Eqs.(2.2.1) and (2.2.2). \qquad $ \Box $  \\ 
 
It is easily seen from Eq.(2.2.1) that for all $ a, \ b \in {\cal R}_{2,0},
 \ \lambda \in {\cal R}$, the following operation properties hold

{\rm (a)} \  $ a=b \ \Longleftrightarrow  \ \phi_{2,0}(a) = \phi_{2,0}(b).$ 

{\rm (b)} \ $ \phi_{2,0}(a + b ) =  \phi_{2,0}(a) + \phi_{2,0}(b), \ \ \
\phi_{2,0}(ab)= \phi_{2,0}(a) \phi_{2,0}(b), \ \ \  \phi_{2,0}( \lambda a)
= \lambda \phi_{2,0}(a).$

{\rm (c)}  \  $ \phi_{2,0}(1)= I_2.$ 

{\rm (d)}  \ $ a = \frac{1}{4}[ \, 1 + e_1, \ e_2 - e_{12} \,]
\phi_{2,0}(a)[  \, 1 + e_1, \ e_2 - e_{12}\, ]^T .$ 

{\rm (e)}  \ $ {\rm det}[\phi_{2,0}(a)] =  a_0^2 - a_1^2 - a_2^2 + a_3^2 ,$ \ for all $  a = a_0 + a_1e_1 +  a_2e_2 + 
 a_3e_{12} \in {\cal R}_{2,0}.$ 

{\rm (f)}  \ $ a $ is invertible $ \Longleftrightarrow \phi_{2,0}(a) $ is
invertible, in which case,  $ \phi_{2,0}(a^{-1}) = \phi_{2,0}^{-1}(a).$ 

{\rm (g)}  \ $ p_a(a ) = 0,$ where $ p_a( x ) = {\rm det}
[x I_2 - \phi_{2,0}(a)].$

{\rm (h)} \ $ a $ is similar to $ b$ over $ {\cal R}_{0,2}$, i.e., there is
an invertible $ x \in {\cal R}_{0,2} $  such that $ xax^{-1} = b $ if and
only if $  \phi_{2,0}(a)$  and  $ \phi_{2,0}(b)$ are similar over
$ {\cal R}$. \\

These properties show that through the bijective map $ \phi_{2,0}: a \in{\cal R}_{2,0} \longrightarrow \phi_{2,0}(a) \in {\cal R}^{2 \times 2},$ the Clifford algebra  ${\cal R}_{2,0}$ is algebraically isomorphic to the matrix algebra $ {\cal R}^{2 \times 2}$, and  $ \phi_{2,0}(a)$ is the matrix representation of $ a $ in
 $ {\cal R}^{2 \times 2}$. \\

\noindent {\bf Theorem 2.2.2.} \ {\em Let $ a \in {\cal R}_{1,1}
= {\cal R} \{ e_1, \ \varepsilon_1 \ | \ e_1^2=1, \ \varepsilon_1^2 = -1 \},$
 the split quaternion algebra. Then $a $ can factor as
 $ a= a_0 + a_1e_1 + a_2 \varepsilon_1 + a_3e_1\varepsilon_1,$ where
 $a_0$---$a_3 \in {\cal R},$ and $ aI_2$  satisfies the following universal
 similarity factorization  equality
$$
 P_{1,1} \left[ \begin{array}{cc} a & 0 \\ 0 & a \end{array} \right] P_{1,1}^{-1}
= \left[ \begin{array}{cc}  a_0 + a_1 &  -(a_2 + a_3) \\   a_2 - a_3 & a_0 - a_1 \end{array} \right] \equiv \phi_{1,1}(a) \in {\cal R}^{2 \times 2 },   \eqno (2.2.5)
$$
where $P_{1,1}$  has the independent form  
$$ 
P_{1,1} = P_{1,1}^{-1} = \frac{1}{2} \left[ \begin{array}{cc} 1+e_1  & \varepsilon_1 - e_1\varepsilon_1 \\    -(\varepsilon_1 + e_1\varepsilon_1)  &  1 - e_1 \end{array} \right].  \eqno (2.2.6) 
$$ } 
 
\noindent {\bf Proof.} \ By Lemma 1.1, we take the change of basis of ${\cal R}_{1,1}$ as follows 
$$ 
\tau_{11}= \frac{1}{2}( 1+e_1), \ \ \ \tau_{12}= - \frac{1}{2}(\varepsilon_1 + e_1\varepsilon_1), \ \ \  
\tau_{21}= \frac{1}{2}(\varepsilon_1 - e_1\varepsilon_1), \ \ \ \tau_{22}= \frac{1}{2}( 1 - e_1). 
 \eqno (2.2.7) 
$$
Then it is not difficult to verify that this new basis satisfies the multiplication laws in (1.10). In that case, every $ a = a_0 + a_1e_1 + a_2 \varepsilon_1 + a_3e_1\varepsilon_1, \in {\cal R}_{1,1}$ can be rewritten as 
$$
a= ( a_0 + a_1)\tau_{11} - ( a_2 + a_3)\tau_{12} + ( a_2 - a_3)\tau_{21} + ( a_0 - a_1)\tau_{22}.
 \eqno (2.2.8)
$$
Substituting Eqs.(2.2.7) and (2.2.8) into Eqs.(1.11) and (1.12),  we  directly obtain Eqs.(2.2.5) and (2.2.6). \qquad  $ \Box $  \\ 
 
Similarly it is easy to verify that through the bijective map $\phi_{1,1}:a \in{\cal R}_{1,1} \longrightarrow \phi_{1,1}(a) \in {\cal R}^{2 \times 2}, $ the Clifford algebra ${\cal R}_{1,1} $ is algebraically isomorphic to the matrix
 algebra $ {\cal R}^{2 \times 2}$, and  $ \phi_{1,1}(a)$ is the matrix representation of $ a $ in $ {\cal R}^{2 \times 2}$. \\

As to the Clifford algebra ${\cal R}_{0,2} = {\cal H}$, the ordinary quaternion division algebra, we have the following two results. \\

\noindent {\bf Theorem 2.2.3.} \ {\em Let $ a=  a_0 + a_1 \varepsilon_1 +
a_2 \varepsilon_2 + a_3 \varepsilon_{12}  \in {\cal H }
= {\cal R}\{ \varepsilon_1 , \ \varepsilon_2 \ | \ \varepsilon_1^2 = -1,
\ \varepsilon_2^2 =-1 \},$ where $ a_0$---$a_3 \in {\cal R}.$
Then $ aI_2$ satisfies the  following universal similarity factorization
equality
$$ 
P_{0,2}\left[ \begin{array}{rr}  a  & 0  \\ 0 &  a \end{array}  \right] P^{-1}_{0,2} 
= \left[ \begin{array}{cc}  a_0 +a_1\varepsilon_1 &  -(a_2 +a_3\varepsilon_1) \\ a_2 - a_3 \varepsilon_1 &  a_0 - a_1\varepsilon_1 \end{array} \right] \equiv \phi_{0,2}(a) \in {\cal C }^{2 \times 2 },  \eqno (2.2.9)
 $$ 
where $P_{0,2}$  has the independent form
$$
P_{0,2} =  P^{-1}_{0,2} = \frac{1}{ \sqrt{2}}\left[ \begin{array}{rr}  1  & -\varepsilon_1  \\ -\varepsilon_2 & \varepsilon_{12} \end{array} \right]. \eqno (2.2.10) 
$$ }

\noindent {\bf Proof.} \ Note that $ a - \varepsilon_1 a \varepsilon_1 = 2( a_0 + a_1\varepsilon_1)$ and $a +  \varepsilon_1 a \varepsilon_1 = 2( a_2\varepsilon_2 + a_3 \varepsilon_{12}).$  We find 
\begin{eqnarray*}
P_{0,2}\left[ \begin{array}{rr}  a  & 0  \\ 0 &  a \end{array}  \right] P^{-1}_{0,2} 
 & = & \frac{1}{2} \left[ \begin{array}{cc}  a - \varepsilon_1 a \varepsilon_1  &  a\varepsilon_2 + \varepsilon_1a \varepsilon_{12} \\ - \varepsilon_2a  + \varepsilon_{12}a\varepsilon_1 &  -(\varepsilon_2a \varepsilon_2 + \varepsilon_{12}a \varepsilon_{12} ) \end{array} \right ] \\
&  = & \frac{1}{2} \left[ \begin{array}{cc}  a - \varepsilon_1 a \varepsilon_1  & ( a + \varepsilon_1 a \varepsilon_1)\varepsilon_2  \\ -\varepsilon_2( a + \varepsilon_1 a \varepsilon_1 ) &  - \varepsilon_2 (a - \varepsilon_1 a \varepsilon_1 )\varepsilon_2 \end{array} \right] \\
& = & \left[ \begin{array}{cc}  a_0 + a_1\varepsilon_1  &  (a_2\varepsilon_2 + a_3 \varepsilon_{12})\varepsilon_2  \\ -\varepsilon_2(a_2\varepsilon_2 + a_3 \varepsilon_{12}) &  - \varepsilon_2(a_0 + a_1\varepsilon_1)\varepsilon_2  \end{array} \right]\\
 & = &  \left[ \begin{array}{cc} a_0 +a_1\varepsilon_1 &  -(a_2 +a_3\varepsilon_1) \\ a_2 - a_3 \varepsilon_1 &  a_0 - a_1\varepsilon_1 \end{array} \right],
\end{eqnarray*} 
which is exactly the desired result.  \qquad   $\Box $ \\

\noindent {\bf Theorem 2.2.4.} \ {\em Let $ a=  a_0 + a_1 \varepsilon_1 +
a_2 \varepsilon_2 + a_3\varepsilon_{12} \in {\cal H},$ where $a_0$---$a_3 \in
 {\cal R} $. Then $ aI_4$ satisfies the following universal similarity
 factorization equality
$$
 Q_{0,2} \left[ \begin{array}{cccc} a &  &  &  \\  & a &  & \\
 &  &  a &    \\  & & & a  \end{array} \right] Q_{0,2}^{-1} = \left[ \begin{array}{rrrr} a_0 & -a_1 & -a_2 
& -a_3  \\ a_1 & a_0 &  -a_3 & a_2 \\ a_2 & a_3 &  a_0 & -a_1  \\ a_3 & -a_2 &  a_1 & a_0  \end{array} \right] \equiv \varphi_{0,2}(a) \in {\cal R }^{4 \times 4 },  \eqno (2.2.11)
$$ 
where $ Q_{0,2} $ has  the independent form 
$$
 Q_{0,2} =  Q_{0,2}^{-1} = \frac{1}{2} \left[ \begin{array}{cccc} 1 & \varepsilon_1  & \varepsilon_2 &
 \varepsilon_{12} \\ -\varepsilon_{1} & 1 & \varepsilon_{12}  & -\varepsilon_{2} \\ - \varepsilon_{2} & 
-\varepsilon_{12} & 1 & \varepsilon_{1}  \\  -\varepsilon_{12} & \varepsilon_{2} & -\varepsilon_{1} &
 1 \end{array} \right].  \eqno (2.2.12)  
$$  }

\noindent {\bf Proof.} \ It is easy to verify that 
$$
  \frac{1}{2} \left[ \begin{array}{rr}  1  &  1 \\ 1 &  -1  \end{array} \right] \left[ \begin{array}{cc} 
 x_0  &  x_1 \\ x_1 &  x_0   \end{array} \right]\left[ \begin{array}{rr}  \ 1 & 1 \\ 1 & -1  \end{array} \right] = \left[ \begin{array}{cc}  x_0 + x_1   &  0 \\  0 &  x_0 - x_1  \end{array} \right]  \eqno (2.2.13)
$$  
holds for any $ x_0, \ x_1 \in {\cal H}$.  On the basis of (2.2.13), we further obtain
$$
\left[ \begin{array}{cccc} x_0 & x_1 & x_2  &  x_3 \\  x_1 & x_0 & x_3  &  x_2 \\ x_2 & x_3 & x_0  &  x_1
 \\ x_3 & x_2 & x_1  &  x_0  \end{array} \right] = V \left[ \begin{array}{cccc} d_1 &  & &  \\
  & d_2 &  &  \\  &  & d_3 &   \\  & &  & d_4  \end{array} \right]V,  \eqno (2.2.14) 
$$ 
where $ x_0, \  x_1, \  x_2, \ x_3 \in {\cal H},$ and 
$$ 
d_1 =  x_0 + x_1 + x_2 + x_3, \qquad d_2 =  x_0 + x_1 - x_2 - x_3,
$$
$$ 
d_3 =  x_0 - x_1 + x_2 - x_3,  \qquad d_4 =  x_0 - x_1 - x_2 + x_3,  
$$ 
$$
V= V^T = V^{-1} = \frac{1}{2} \left[ \begin{array}{rrrr} 1 & 1 & 1 & 1 \\ 1 & 1 & -1 & -1 
\\ 1 & -1 & 1  & -1  \\ 1 &  -1 & -1 & 1 \end{array} \right].  
$$ 
Replacing $x_0$---$x_3$ in Eq.(2.2.14) now by $ a_0, \ a_1\varepsilon_1, \ a_2\varepsilon_2$ and $a_3\varepsilon_{12}$, respectively, we obtain 
$$
A \equiv \left[ \begin{array}{cccc} a_0 & a_1\varepsilon_1  & a_2\varepsilon_2  & a_3 \varepsilon_{12} \\  a_1\varepsilon_1 & a_0 & a_3\varepsilon_{12}  & a_2\varepsilon_2 
 \\ a_2\varepsilon_2 &  a_3\varepsilon_{12} & a_0  & a_1\varepsilon_1 \\  a_3\varepsilon_{12} & a_2\varepsilon_2 & a_1\varepsilon_1  & a_0  \end{array} \right] = 
V\left[ \begin{array}{cccc}  d_1 & & &  \\  & d_2 &
 &  \\ & & d_3 & \\ & & & d_4 \end{array} \right]V,   \eqno (2.2.15)
$$
where
 $$ 
d_1 =  a_0 + a_1\varepsilon_1 + a_2\varepsilon_2 + a_3\varepsilon_{12} = a, \qquad  
d_2 =  a_0 + a_1\varepsilon_1 - a_2\varepsilon_2 - a_3\varepsilon_{12},  
$$ 
$$ 
d_3 =  a_0 - a_1\varepsilon_1 + a_2\varepsilon_2 - a_3\varepsilon_{12}, \qquad d_4 =  a_0 - a_1\varepsilon_1 - a_2\varepsilon_2 + a_3\varepsilon_{12}. 
$$ 
It is easy to verify that $  d_2 = \varepsilon_1a\varepsilon_1^{-1}, \ d_3 = \varepsilon_2a\varepsilon_2^{-1}$ and $ d_4 = \varepsilon_{12}a\varepsilon_{12}^{-1}.$ Thus
$$ 
{\rm diag}( \, d_1, \ d_2, \  d_3, \  d_4 \, ) = J{\rm diag}( \, a, \ a, \ a, \ a \,)J^{-1}, 
\eqno (2.2.16)
$$
where $J = {\rm diag}( \, 1 , \ \varepsilon_1, \ \varepsilon_2, \ \varepsilon_{12} \, ).$ On the other hand, it is easy to verify that 
$$
J^{-1}AJ = \left[ \begin{array}{cccc} a_0 & a_1\varepsilon_1\varepsilon_1  & a_2\varepsilon_2\varepsilon_2 & a_3\varepsilon_{12}\varepsilon_{12}  
\\  a_1\varepsilon_{1}^{-1}\varepsilon_{1} & a_0\varepsilon_{1}^{-1}\varepsilon_{1} & a_3\varepsilon_{1}^{-1}\varepsilon_{12}\varepsilon_{2} & a_2\varepsilon_{1}^{-1}\varepsilon_{2}\varepsilon_{12}  \\
 a_2\varepsilon_{2}^{-1}\varepsilon_{2} & a_3\varepsilon_{2}^{-1}\varepsilon_{12}\varepsilon_{1} & a_0\varepsilon_{2}^{-1}\varepsilon_{2}  & a_1\varepsilon_{2}^{-1}\varepsilon_{1}\varepsilon_{12} \\
 a_3\varepsilon_{12}^{-1}\varepsilon_{12} & a_2\varepsilon_{12}^{-1}\varepsilon_{2}\varepsilon_{1} &
 a_1\varepsilon_{12}^{-1} \varepsilon_{1}\varepsilon_{2} & a_0 \varepsilon_{12}^{-1}\varepsilon_{12} \end{array} \right] = \left[ \begin{array}{rrrr} a_0 & -a_1 &  -a_2 & -a_3  \\ a_1 & a_0 &  -a_3 & a_2 \\
a_2 & a_3 &  a_0 &  -a_1  \\ a_3 & -a_2 &  a_1 & a_0  \end{array} \right] = \phi(a). 
$$
Putting Eqs.(2.2.15) and (2.2.16) in it yields
$$
 \phi(a) = J^{-1}AJ = J^{-1}V{\rm diag}( \, d_1, \ d_2, \  d_3, \  d_4 \, )VJ = (J^{-1}VJ){\rm diag}( \, a, \ a, \ a, \ a \,)(J^{-1}VJ).
$$
Let $ P = J^{-1}VJ.$ Then we have Eqs.(2.2.11) and (2.2.12).  \qquad  $ \Box $ \\

Just as the results in Theorems 2.1.3 and 2.1.4, the universal similarity factorization equalities in Theorems 2.2.1---2.2.4 can also be  extend to all matrices over $ {\cal R}_{2,0}, \ {\cal R}_{1,1}$ and $ {\cal R}_{0,2}$, respectively. We leave them to the reader. \\

\noindent {\bf 2.3. The Cases for ${\cal R}_{p,q}$ with $ p +q = 3$ } \\

According to the multiplication laws in Eqs.(1.2) and (1.3) we know that for all algebras ${\cal R}_{p,q}$ with
 $ p + q = n $ odd, the commutative rule 
$$ 
ae_1e_2 \cdots e_n =  e_1e_2 \cdots e_na 
$$ 
holds for all $ a \in {\cal R}_{p,q}$. Based on this simple fact, we can write all elements of
 ${\cal R}_{p,q}$ with $ p + q = 3$ in the form 
$$ 
a = a_0 + a_1e_{[3]},  \eqno (2.3.1)
$$ 
where $ a_0, \ a_1 \in{\cal R}\{ e_1, \ e_2 \},$ or $ a_0, \ a_1 \in{\cal R}\{ e_1, \ e_3 \},$  or  $ a_0, \ a_1 \in{\cal R}\{ e_2, \ e_3 \}$. From  (2.3.1) we can introduce the conjugate of $ a $ as follows
$$
 \overline{a} =  a_0 - a_1e_{123}. \eqno (2.3.2)
$$
Then it is easy to verify that for all $ a, \ b \in {\cal R}_{p,q}$ and
$ \lambda \in {\cal R}$,
$$ 
\overline{\overline{a}} = a,  \qquad    \overline{ a + b } = \overline{ a } + \overline{ b }, \qquad   
\overline{ab}=\overline{a}\overline{b}, \qquad    \overline {\lambda a} = \overline { a \lambda } = 
 \lambda \overline{a}.
$$ 
In this subsection we combine Eqs.(2.3.1) and (2.3.2) with the results in Subsection 2.2 to establish four 
universal similarity factorization equalities over ${\cal R}_{p,q}$ with $ p + q = 3$.  \\

\noindent {\bf Theorem 2.3.1.} \ {\em Let $ a \in {\cal R}_{3,0} =
 {\cal R}\{\, e_1, \ e_2, \ e_3 \, \},$ and write $a$ as  $ a=  a_0 + a_1e_{[3]},$ where
$$
a_0, \ a_1 \in {\cal R}_{2,0} = {\cal R}\{ \, e_1, \ e_2 \, \}, \qquad \ e_{[3]}^2  = -1.
$$
Then $ aI_2$  satisfies the following universal similarity factorization
 equality
$$
 P_{3,0} \left[ \begin{array}{cc}  a  & 0   \\ 0 & a  \end{array} \right] P_{3,0}^{-1}
=  \phi_{2,0}(a_0) + \phi_{2,0}(a_1) e_{[3]} \equiv \phi_{3,0}(a) \in {\cal C}^{2 \times 2},  
\eqno (2.3.3)
$$
where\ $ \phi_{2,0}(a_t)( t = 0, \ 1 )$ is the matrix representation of $ a_t$ in $ {\cal R}^{2 \times 2}$ 
 defined in Eq.{\rm (2.2.1)} and  $P_{3,0}$ has the independent form 
$$
P_{3,0}  = P_{3,0}^{-1} =  P_{2,0} = \frac{1}{2} \left[ \begin{array}{cc} 1 + e_1  & e_2 - e_{12} \\  e_2 + e_{12} &  1 - e_1 \end{array} \right].  \eqno (2.3.4)
$$ } 

\noindent  {\bf Proof.} \ Writing $ aI_2$ as $ aI_2 = a_0 I_2 + a_1e_{[3]}I_2$ and multiplying $  P_{2,0}$ and  $ P_{2,0}^{-1}$  on its both sides, we obtain 
\begin{eqnarray*}
P_{2,0}(aI_2) P_{2,0}^{-1} & = & P_{2,0}(a_0I_2) P_{2,0}^{-1} +  P_{2,0}(a_1e_{[3]}I_2) P_{2,0}^{-1} \\
& = & P_{2,0}(a_0I_2) P_{2,0}^{-1} +  P_{2,0}(a_1I_2) P_{2,0}^{-1}e_{[3]}\\ 
& = & \phi_{2,0}(a_0) + \phi_{2,0}(a_1) e_{[3]} =  \phi_{3,0}(a).
\end{eqnarray*}
Note that $ \phi_{2,0}(a_t) \in {\cal R}^{2 \times 2}$ and $ e_{[3]}^2  = -1.$ Thus (2.3.3) follows. 
\qquad $ \Box $ \\

Obviously, through the bijective map $ \phi_{3,0}: a \in{\cal R}_{3,0} \longrightarrow \phi_{3,0}(a) \in {\cal C}^{2 \times 2}, $ the Clifford algebra  ${\cal R}_{3,0}$ is algebraically isomorphic to the matrix algebra $ {\cal C}^{2 \times 2}$, and  $ \phi_{3,0}(a)$ is the matrix representation of $ a $ in 
$ {\cal C}^{2 \times 2}$. \\

\noindent {\bf Theorem 2.3.2.} \ {\em Let $ a \in {\cal R}_{2,1} =
{\cal R} \{ e_1, \ e_2, \  \varepsilon_1 \, \},$ and write $a$ as
$a=  a_0 + a_1e,$ where
$$
a_0, \ a_1 \in {\cal R}_{1,1} = {\cal R}\{ e_1, \  \varepsilon_1 \, \}, \qquad
 e = e_1e_2\varepsilon_1,   \qquad  e^2  = 1.
$$
Then the diagonal matrix $ D_a = {\rm diag}( aI_2 , \  \overline{a}I_2 ) $ satisfies the following universal  similarity factorization  equality 
$$
 P_{2,1}D_a P_{2,1}^{-1}
 = \left[ \begin{array}{cc} \phi_{1,1}(a_0) + \phi_{1,1}(a_1) & O  \\ O   & \phi_{1,1}(a_0) - \phi_{1,1}(a_1) \end{array} \right] \equiv \phi_{2,1}(a) \in \, ^2{\cal R }^{ 2 \times 2},   \eqno (2.3.5)
 $$ 
where\ $ \phi_{1,1}(a_t)( t = 0, \ 1 )$ is the matrix representation of $ a_t$  in  $ {\cal R}^{2 \times 2}$ defined  in Eq.{\rm (2.2.5)}$,$ and 
$$
P_{2,1} = \frac{1}{2} \left[ \begin{array}{cc} (1 + e)P_{1,1} &  -( 1 - e)P_{1,1} \\  (1 - e)P_{1,1} & 
 (1 + e)P_{1,1} \end{array} \right], \ \ \
 P_{2,1}^{-1} = \frac{1}{2} \left[ \begin{array}{cc} P_{1,1}^{-1}(1 + e) & P_{1,1}^{-1}( 1 - e) \\
  -P_{1,1}^{-1}(1 - e) & P_{1,1}^{-1}(1 + e) \end{array} \right],   \eqno (2.3.6)
$$
$ P_{1,1} $ and $ P_{1,1}^{-1} $ are given by Eq.{\rm (2.2.6)}. } \\ 
  
\noindent {\bf Proof.} \ Writing $ aI_2$ as $ aI_2 = a_0 I_2 + a_1eI_2$ and multiplying 
$ P_{1,1}$ and  $ P_{1,1}^{-1}$ in Eq.(2.2.6)  on its both sides, we get 
\begin{eqnarray*}
P_{1,1}(aI_2) P_{1,1}^{-1} 
& = & P_{1,1}(a_0I_2) P_{1,1}^{-1} +  P_{1,1}(a_1eI_2) P_{1,1}^{-1} \\
& = & P_{1,1}(a_0I_2) P_{1,1}^{-1} +  P_{1,1}(a_1I_2) P_{1,1}^{-1}e = \phi_{1,1}(a_0) + \phi_{1,1}(a_1) e 
=  \psi(a),
\end{eqnarray*}
where $  \phi_{1,1}(a_t) \in {\cal R}^{2 \times 2}( t = 0, \ 1 ).$ Here we denote $\overline{\psi}(a) = \phi_{1,1}(a_0)  - \phi_{1,1}(a_1)e.$  Then $ \psi(\overline{a} ) = \overline{\psi}(a)$. Note that $ e^2  = 1.$ Thus by Theorem 2.2.1, we can build a matrix and its inverse as follows 
$$ 
V =  \frac{1}{2} \left[ \begin{array}{cc} (1 + e)I_2 &  -( 1 - e)I_2 \\  (1 - e)I_2 & 
 (1 + e) I_2\end{array} \right], \qquad V^{-1} = \frac{1}{2} \left[ \begin{array}{cc}(1 + e)I_2 &
 ( 1 - e)I_2 \\  -(1 - e)I_2 &  (1 + e)I_2 \end{array} \right].
$$
Now applying them to diag$( \, \psi(a), \  \psi(\overline{a}) \, )$ we obtain
\begin{eqnarray*}
V \left[ \begin{array}{cc} \psi(a)  &  O \\ O &  \psi(\overline{a}) \end{array} \right] V^{-1} 
& = & V \left[ \begin{array}{cc}  \phi_{1,1}(a_0) + \phi_{1,1}(a_1)e  & O \\ O   & \phi_{1,1}(a_0) +
 \phi_{1,1}(a_1)e \end{array} \right] V^{-1} \\
&  = &  \left[ \begin{array}{cc}  \phi_{1,1}(a_0) + \phi_{1,1}(a_1) & O  \\ O   & \phi_{1,1}(a_0) -
 \phi_{1,1}(a_1) \end{array} \right].
\end{eqnarray*}
Finally substituting $ \psi(a) = P_{1,1}(aI_2) P_{1,1}^{-1}$ and  $ \psi(\overline{a}) = 
P_{1,1}(\overline{a}I_2)P_{1,1}^{-1}$ into the left-hand side of the  above equality yields Eq.(2.3.5). \qquad   $ \Box $ \\

\noindent {\bf Theorem 2.3.3.} \ {\em Let $ a \in {\cal R}_{1,2}
= {\cal R}\{ e_1, \ \varepsilon_1, \ \varepsilon_2 \, \},$ and write $a$ as
$  a=  a_0 + a_1e,$ where
$$
a_0, \ a_1 \in {\cal R}_{1,1} = {\cal R}\{ e_1, \ \varepsilon_1 \, \}, \qquad e = e_1\varepsilon_1\varepsilon_2 , \qquad  e^2  = -1.
$$
Then $ aI_2$ satisfies the following universal similarity  factorization
equality
$$
 P_{1,2} \left[ \begin{array}{cc}  a  & 0   \\ 0 & a  \end{array} \right] P_{1,2}^{-1}
=  \phi_{1,1}(a_0) + \phi_{1,1}(a_1)e \equiv \phi_{1,2}(a) \in {\cal C}^{2 \times 2},  \eqno (2.3.7)
$$
where\ $ \phi_{1,1}(a_t)( t = 0, \ 1 )$ is the matrix representation of $ a_t$ in 
$ {\cal R}^{2 \times 2}$ defined in Eq.{\rm (2.2.5)}$,$ and 
$$
P_{1,2}  = P_{1,2}^{-1}  = P_{1,1} = \frac{1}{2} \left[ \begin{array}{cc} 1+e_1  & \varepsilon_1 - e_1\varepsilon_1 \\    -(\varepsilon_1 + e_1\varepsilon_1)  &  1 - e_1 \end{array} \right]. 
 \eqno (2.3.8) 
$$ }

\noindent {\bf Proof.} \ Writing $ aI_2 = a_0 I_2 + a_1eI_2$ and multiplying $  P_{1,1}$ 
and  $ P_{1,1}^{-1}$  on its both sides, we get 
\begin{eqnarray*}
P_{1,1}(aI_2) P_{1,1}^{-1} & = & P_{1,1}(a_0I_2) P_{1,1}^{-1} +  P_{1,1}(a_1I_2) P_{1,1}^{-1} \\
& = & P_{1,1}(a_0I_2) P_{1,1}^{-1} +  P_{1,1}(a_1I_2) P_{1,1}^{-1}e  =  \phi_{1,1}(a_0) + \phi_{1,1}(a_1)e. 
\end{eqnarray*}
Note that $ \phi_{1,1}(a_t) \in {\cal R}^{2 \times 2}$ and $ e^2  = -1.$ Thus Eq.(2.3.7) follows. 
 \qquad $ \Box $ \\

\noindent {\bf Theorem 2.3.4.} \ {\em Let $ a \in {\cal R}_{0,3}
= {\cal R}\{ \, \varepsilon_1, \ \varepsilon_2, \
 \varepsilon_3 \, \},$ and write $a$ as $ a=  a_0 + a_1\varepsilon_{[3]},$
 where
$$
a_0, \ a_1 \in {\cal R}_{0,2} = {\cal R}\{ \, \varepsilon_1, \  \varepsilon_2
 \, \} = {\cal H }, \qquad \varepsilon^2_{[3]}  = 1.
$$
Then $ a $  and $ \overline{a} = a_0 - a_1\varepsilon_{[3]}$  satisfy the following universal similarity factorization equality 
$$
 P_{0,3} \left[ \begin{array}{cc}  a &  0  \\ 0 & \overline{a} \end{array} \right] P_{0,3}^{-1}
 = \left[ \begin{array}{cc} a_0 + a_1 & 0  \\ 0   & a_0 - a_1  \end{array} \right] \equiv \phi_{0,3}(a) 
\in \, ^2{\cal H },   \eqno (2.3.9)
 $$ 
where
$$
P_{0,3} = \frac{1}{2} \left[ \begin{array}{cc} 1 + \varepsilon_{[3]} &  -( 1 - \varepsilon_{[3]}) \\ 1 - \varepsilon_{[3]} &  1 + \varepsilon_{[3]} \end{array} \right], \qquad P_{0,3}^{-1} = \frac{1}{2} \left[ \begin{array}{cc}  1 + \varepsilon_{[3]} &  1 - \varepsilon_{[3]} \\ -(1 - \varepsilon_{[3]}) &  1 + 
\varepsilon_{[3]} \end{array} \right].   \eqno (2.3.10)
$$ } 

The derivation of Eq.(2.3.5) is much analogous to that of Eq.(2.1.1). So we omit it here. \\
     
\noindent {\bf 2.4. The Cases for $ {\cal R}_{p,q} $ with  $ p +q = 4$ } \\

The five  algebraic isomorphisms for $ \in {\cal R}_{p,q} $ with $ p + q = 4$ are shown in Eq.(1.7).  Without much effort we can extend the results in Subsection 2.2 to these five algebras.  \\

\noindent {\bf Theorem 2.4.1.} \ {\em Let $ a \in {\cal R}_{4,0}
= {\cal R}\{ \, e_1, \ e_2, \ e_3, \ e_4 \, \}.$
Then $ a $ can factor as    
$$
a = a_0 + a_1e_{123} + a_2e_{124} + a_3e_{34}, 
$$ 
where 
$$
a_0, \ a_1, \ a_2, \ a_3  \in {\cal R}_{2,0} = {\cal R}\{ e_1, \ e_2 \, \},
$$
$$ 
e_{123}^2  = -1, \qquad  e_{124}^2  = -1, \qquad
e_{34}= -e_{123}e_{124} = e_{124}e_{123}.
$$
In that case $ aI_2$ satisfies the following universal similarity factorization equality
$$
 P_{4,0} \left[ \begin{array}{cc}  a  & 0   \\ 0 & a  \end{array} \right] P_{4,0}^{-1}
=  \phi_{2,0}(a_0) +  \phi_{2,0}(a_1)e_{123}  +  \phi_{2,0}(a_2)e_{124} +  \phi_{2,0}(a_3)e_{34} 
 \equiv \phi_{4,0}(a),   \eqno (2.4.1) 
$$
where
$$
\phi_{4,0}(a) \in  {\cal R}^{2\times 2} \{ \, e_{123}, \ e_{124}  \, \} =  {\cal H}^{2 \times 2},  \eqno (2.4.2)
$$
$ \phi_{2,0}(a_t)( t = 0$---$3)$ is the matrix representation of $ a_t$ in
$ {\cal R}^{2 \times 2}$ defined in Eq.{\rm (2.2.1)}$,$  and
$ P_{4,0}  = P_{4,0}^{-1} =  P_{2,0}, $ the matrix in Eq.{\rm (2.2.2)}.  } \\

\noindent {\bf Proof.} \ Note that  the commutative rules
 $ be_{[3]} = e_{[3]}b, \ be_{124} = e_{124}b$ and $be_{34} = e_{34}b $ hold
 for all $ b \in {\cal R}_{2,0} = {\cal R}\{ \, e_1, \ e_2 \, \}.$ Thus we
 immediately obtain
\begin{eqnarray*}
\lefteqn{ P_{2,0}(aI_2) P_{2,0}^{-1} } \\
& =& P_{2,0}(a_0I_2)P_{2,0}^{-1} + P_{2,0}(a_1I_2) P_{2,0}^{-1}e_{[3]} 
+ P_{2,0}(a_2I_2) P_{2,0}^{-1}e_{124} + P_{2,0}(a_3I_2) P_{2,0}^{-1}e_{34} \\
& = & \phi_{2,0}(a_0) + \phi_{2,0}(a_1)e_{123} + \phi_{2,0}(a_2)e_{124} + \phi_{2,0}(a_3)e_{34},
\end{eqnarray*}
which is exactly the result in Eq.(2.4.1).   \qquad  $ \Box $ \\   

\noindent {\bf Theorem 2.4.2.} \ { \em Let $ a \in {\cal R}_{3,1}
= {\cal R}\{ \, e_1, \ e_2, \ e_3, \ \varepsilon_1 \,\}.$
Then $a$ can factor as    
$$
a = a_0 + a_1(e_{12}\varepsilon_1) + a_2e_{[3]}+ a_3(e_3\varepsilon_1),  
$$ 
where 
$$
a_0, \ a_1, \ a_2, \ a_3 \in {\cal R}_{2,0} = {\cal R}\{\, e_1, \ e_2 \, \},
$$
$$ 
 (e_{12}\varepsilon_1)^2 = 1, \qquad  e_{[3]}^2  = -1, \qquad e_3\varepsilon_1
 = (e_{12}\varepsilon_1)e_{[3]}= -e_{[3]}(e_{12}\varepsilon_1). 
$$
In that case $ aI_4$  satisfies the following universal similarity factorization  equality 
$$
 P_{3,1}(aI_4) P_{3,1}^{-1}= \left[ \begin{array}{cc}  \phi_{2,0}(a_0) + \phi_{2,0}(a_1)  &  -[ \ \phi_{2,0}(a_2) + \phi_{2,0}(a_3) \ ]
 \\ \phi_{2,0}(a_2) - \phi_{2,0}(a_3)   & \phi_{2,0}(a_0) - \phi_{2,0}(a_1)  \end{array} \right] 
\equiv \phi_{3,1}(a) \in  {\cal R}^{4\times 4}, \eqno (2.4.3)
$$
where $\phi_{2,0}(a_t)( t = 0$---$3)$ is the matrix representation of
 $ a_t$ in ${\cal R}^{2 \times 2}$ defined in Eq.{\rm (2.2.1)} and 
$$
P_{3,1} = P_{3,1}^{-1} = \frac{1}{2} \left[ \begin{array}{cc} ( 1 + e_{12}\varepsilon_1 )P_{2,0}
 &  ( e_{[3]} - e_3\varepsilon_2) P_{2,0} \\ - ( e_{[3]} - e_3\varepsilon_1) P_{2,0}
  & ( 1 + e_{12}\varepsilon_1 )P_{2,0} \end{array} \right],  \eqno (2.4.4)
$$ 
where $ P_{2,0}$ is given in Eq.{\rm (2.2.2)}. } \\

\noindent  {\bf Proof.} \ Note that the following commutative laws
 $be_{[3]}  = e_{[3]} b, \ be_{124} = e_{124}b$ and
  $be_{34} = e_{34}b $ hold for all $ b \in {\cal R}_{2,0}
  = {\cal R}\{\, e_1, \ e_2 \, \}.$ Thus it follows from (2.2.1)
that
\begin{eqnarray*}
\lefteqn{ P_{2,0}(aI_2) P_{2,0}^{-1}} \\
& = & P_{2,0}(a_0I_2)P_{2,0}^{-1} + P_{2,0}(a_1I_2) P_{2,0}^{-1}(e_{12}\varepsilon_1) 
+ P_{2,0}(a_2I_2) P_{2,0}^{-1}e_{[3]} + P_{2,0}(a_3I_2) P_{2,0}^{-1}(e_3\varepsilon_1) \\
& = & \phi_{2,0}(a_0) + \phi_{2,0}(a_1)(e_{12}\varepsilon_1) + \phi_{2,0}(a_2)e_{[3]} + \phi_{2,0}(a_3)
(e_3\varepsilon_1 ) \\
& \equiv & \psi( a ) \in {\cal R}^{ 2\times 2}\{ \, e_{12}\varepsilon_1,
\ e_{[3]} \, \}.
\end{eqnarray*}
Next  building  a matrix  and its inverse from the basis $ 1, \ e_{12}\varepsilon_1, \ e_{[3]} $ and $ e_3\varepsilon_1$ as follows 
$$  
V = V^{-1} = \frac{1}{2} \left[ \begin{array}{cc} (1 + e_{12}\varepsilon_1 )I_2 & 
( e_{[3]} - e_3\varepsilon_1)I_2 \\ -( e_{[3]} + e_3\varepsilon_1)I_2  & (1 - e_{12}\varepsilon_1 )I_2 \end{array} \right],
$$ 
and applying them to the matrix $ \psi( a ) $ given above, we obtain
$$
 V \left[ \begin{array}{cc}  \psi( a )   & O  \\ O &  \psi( a ) \end{array} \right]V^{-1} 
= \left[ \begin{array}{cc}  \phi_{2,0}(a_0) + \phi_{2,0}(a_1)  &  -[ \ \phi_{2,0}(a_2) + \phi_{2,0}(a_3) \ ]
 \\ \phi_{2,0}(a_2) - \phi_{2,0}(a_3)   & \phi_{2,0}(a_0) - \phi_{2,0}(a_1)  \end{array} \right] 
\equiv \phi_{3,1}(a).  
$$
Finally substituting $\psi( a ) = P_{2,0}(aI_2)P_{2,0}^{-1}$ into  the left-hand side of the above equality yields Eqs.(2.4.3) and (2.4.4). \qquad  $ \Box $ \\

Similarly we have the following. \\

\noindent {\bf Theorem 2.4.3.} \ {\em Let $ a \in {\cal R}_{2,2}
= {\cal R} \{  \, e_1, \ e_2, \ \varepsilon_1, \
\varepsilon_2 \, \}, $ the split Clifford algebra. Then  $a$ can factor as   
$$
a = a_0 + a_1(e_{12}\varepsilon_1) + a_2(e_1 \varepsilon_{12}) + a_3(\varepsilon_2e_2),  
$$ 
where 
$$
a_0, \ a_1, \ a_2, \ a_3  \in {\cal R}_{1,1}
= {\cal R}\{ \, e_1, \varepsilon_1 \, \},
$$
$$ 
 (e_{12}\varepsilon_1)^2 = 1, \qquad ( e_1 \varepsilon_{12}) ^2  = -1, \qquad  \varepsilon_2e_2
 = (e_{12}\varepsilon_1)( e_1 \varepsilon_{12})= -( e_1\varepsilon_{12})(e_{12}\varepsilon_1).
$$
In that case $ aI_4$  satisfies the following universal similarity factorization  equality 
$$
 P_{2,2}( aI_4) P_{2,2}^{-1} = \left[ \begin{array}{cc}  \phi_{1,1}(a_0) + \phi_{1,1}(a_1)  & 
 -[\ \phi_{1,1}(a_2) + \phi_{1,1}(a_3) \ ]
 \\ \phi_{1,1}(a_2) - \phi_{1,1}(a_3)   & \phi_{1,1}(a_0) - \phi_{1,1}(a_1) \end{array} \right] 
\equiv \phi_{2,2}(a) \in {\cal R}^{4\times 4}, \eqno (2.4.5)
$$
where $ \phi_{1,1}(a_t)( t = 0$---$3)$ is the matrix representation of $ a_t$ in $ {\cal R}^{2 \times 2}$ 
 given in {\rm (2.2.1)} and 
$$
P_{2,2} = P_{2,2}^{-1} =\frac{1}{2} \left[ \begin{array}{cc} ( 1 + e_{12}\varepsilon_1 )P_{1,1}
 &  ( e_1 \varepsilon_{12} - e_2\varepsilon_2) P_{1,1} \\ - ( e_1 \varepsilon_{12} - e_2\varepsilon_2) P_{1,1}
  & ( 1 + e_{12}\varepsilon_1 )P_{1,1} \end{array} \right],  \eqno (2.4.6)
$$ 
where $ P_{1,1}$ is given in Eq.{\rm (2.2.6)}. } \\

\noindent {\bf Theorem 2.4.4.} \ {\em Let $ a \in {\cal R}_{1,3} =
 {\cal R}\{ \,  e_1, \ \varepsilon_1, \ \varepsilon_2,
  \ \varepsilon_3 \,\}.$
Then $a$ can factor as   
$$
a = a_0 + a_1\varepsilon_{[3]} + a_2e_1( \varepsilon_{12}) + a_3(e_1\varepsilon_3),  
$$ 
where 
$$
a_0, \ a_1, \ a_2, \ a_3 \in {\cal R}_{0,2}
= {\cal R}\{ \, \varepsilon_1, \ \varepsilon_2 \, \} = {\cal H},
$$
$$ 
 \varepsilon_{[3]}^2 = 1, \qquad  ( e_1 \varepsilon_{12})^2  = -1, \qquad
 e_1 \varepsilon_3 = \varepsilon_{[3]}( e_1 \varepsilon_{12})= -( e_1\varepsilon_{12})\varepsilon_
{[3]}. 
$$
In that case $ aI_2$  satisfies the following universal similarity factorization  equality 
$$
 P_{1,3}( aI_2) P_{1,3}^{-1} = \left[ \begin{array}{cc} a_0 + a_1  & -( a_2 +  a_3) \\ a_2 - a_3   &
 a_0 - a_1 \end{array} \right] \equiv \phi_{1,3}(a) \in {\cal H}^{2 \times 2}, \eqno (2.4.7)
$$
where
$$
P_{1,3} = P_{1,3}^{-1} =\frac{1}{2} \left[ \begin{array}{cc}  1 + \varepsilon_{[3]}  & 
e_1\varepsilon_{12} - e_1\varepsilon_3 \\ - ( e_1\varepsilon_{12} - e_1\varepsilon_3)
  & 1 - \varepsilon_{[3]} \end{array} \right]. \eqno (2.4.8)
$$  } 

\noindent {\bf Proof.} \ Note that the following commutative laws
$ b\varepsilon_{[3]} = \varepsilon_{[3]}b, \ b (e_1\varepsilon_{12}) =
 ( e_1\varepsilon_{12} )b,$ $ b(e_1\varepsilon_3) = ( e_1\varepsilon_3)b $
 hold for all $ b \in {\cal R}_{0,2}
 = {\cal R}\{\, \varepsilon_1, \ \varepsilon_2 \,\}.$
Thus by Theorem 2.2.2 we can build the matrix $ P_{1,3} $ in  Eq.(2.4.8) such that $ aI_2$ satisfies
Eq.(2.4.7).  \qquad $\Box $ \\
 
\noindent  {\bf Theorem 2.4.5.} \ {\em Let $ a \in {\cal R}_{0,4}
 = {\cal R}\{\, \varepsilon_1, \ \varepsilon_2,
  \ \varepsilon_3, \ \varepsilon_4 \,\}.$
Then $a$ can factor as   
$$
a = a_0 + a_1\varepsilon_{123} + a_2\varepsilon_{124} + a_3\varepsilon_{43}, 
$$ 
where 
$$
a_0, \ a_1, \ a_2, \ a_3 \in {\cal R}_{0,2}
= {\cal R}\{ \, \varepsilon_1, \ \varepsilon_2 \, \} = {\cal H},
$$
$$ 
 \varepsilon_{123}^2 = 1, \qquad   \varepsilon_{124}^2  = 1,  \qquad \varepsilon_{43}  = 
\varepsilon_{123}\varepsilon_{124} = - \varepsilon_{124}\varepsilon_{123}. 
$$
In that case $ aI_2$  satisfies the following universal similarity factorization  equality 
$$
 P_{0,4}( aI_2) P_{0,4}^{-1} = \left[ \begin{array}{cc} a_0 + a_1  &  a_2 +  a_3 \\ a_2 - a_3 & 
a_0 - a_1 \end{array} \right] \equiv \phi_{0,4}(a) \in {\cal H}^{2\times 2}, \eqno (2.4.9)
$$
where
$$
P_{0,4} = P_{0,4}^{-1} =\frac{1}{2} \left[ \begin{array}{cc}  1 + \varepsilon_{123}
 &  \varepsilon_{124} - \varepsilon_{43} \\ \varepsilon_{124} - \varepsilon_{34}
  & 1 - \varepsilon_{123} \end{array} \right].  \eqno (2.4.10)
$$  } 

\noindent {\bf Proof.} \  Follows from the fact $ b\varepsilon_{123} = \varepsilon_{123}b, \  b\varepsilon_{124}
 = e_1\varepsilon_{124}b, \  b\varepsilon_{43} = \varepsilon_{43}b $ for all
  $ b \in {\cal R}_{0,2} = {\cal R}\{ \, \varepsilon_1, \ \varepsilon_2 \, \}
   $ and Theorem 2.2.1.  \qquad  $\Box $ \\
 
\noindent {\bf 2.5. The Cases for $ {\cal R}_{p,q}$ with $ p+ q= 5$ } \\

Just as for $ {\cal R}_{p,q}$ with $ p+ q= 3 $, we can write all elements of $ {\cal R}_{p,q}$ with
 $ p +q = 5$  in the following form
$$ 
a = a_0 + a_1e_{[5]} = a_0 + e_{[5]}a_1, 
$$ 
where $ a_0, \ a_1$ are the elements of the Clifford algebras defined on any  four generators of $ e_1, \  e_2,
\ \cdots, \  e_5$.  Besides we can  also introduce the conjugate of $ a $ as follows
$$
 \overline{a} =  a_0 - a_1e_{[5]}.
$$
Then it is easy to verify that for all $ a, \ b \in {\cal R}_{p,q}$ and
$ \lambda \in {\cal R}$,
$$ 
\overline{\overline{a}} = a, \qquad \overline{ a + b } = \overline{ a } + \overline{ b }, \qquad
\overline{ab}=\overline{a}\overline{b}, \qquad \overline {\lambda a} = \overline { a \lambda } = 
 \lambda \overline{a}.
$$

According to the table in Eq.(1.4), we know that the six Clifford algebras ${\cal R}_{p,q}$ with $ p+ q= 5 $ 
satisfy the following algebraic isomorphisms 
$$ 
{\cal R}_{5,0} \  \simeq   \ ^2{\cal H}^{2 \times 2}, \qquad 
{\cal R}_{4,1} \  \simeq  \ {\cal C}^{4 \times 4}, \qquad
{\cal R}_{3,2} \  \simeq  \  ^2{\cal R}^{4 \times 4},   \eqno (2.5.1)
$$
$$
{\cal R}_{2,3} \  \simeq \ {\cal C}^{4 \times 4},  \qquad
{\cal R}_{1,4} \  \simeq \  ^2{\cal H}^{2 \times 2},  \qquad 
{\cal R}_{0,5} \  \simeq  \ {\cal C}^{4 \times 4}.   \eqno (2.5.2)
$$
Based on the results in previous  subsection we can establish six universal similarity  factorization equalities  between
 elements of  ${\cal R}_{p,q}$ with $ p + q = 5$ and  matrices of the  six matrix algebras in Eqs.(2.5.1) and 
(2.5.2).\\ 

\noindent  {\bf Theorem 2.5.1.} \ {\em Let $ a \in {\cal R}_{5,0}
= {\cal R}\{ e_1, \ \cdots, \ e_5  \},$ and write $a$ as $ a=  a_0 + a_1e_{[5]},$ where
$$
a_0, \ a_1 \in {\cal R}_{4,0} = {\cal R} \{ \, e_1, \ e_2, \ e_3,  \ e_4 \, \},
\qquad e_{[5]}^2  = 1.  
$$
Then  the diagonal matrix $ D_a = {\rm diag} ( aI_2 , \  \overline{a}I_2 ) $
satisfies the following  universal similarity  factorization equality
$$
 P_{5,0}D_a P_{5,0}^{-1}
 = \left[ \begin{array}{cc} \phi_{4,0}(a_0) + \phi_{4,0}(a_1) & O  \\ O   & \phi_{4,0}(a_0) - \phi_{4,0}(a_1) \end{array} \right] \equiv \phi_{5,0}(a) \in \, ^2{\cal H }^{ 2 \times 2},   \eqno (2.5.3)
$$ 
where  $ \phi_{4,0}(a_t)( t = 0, \ 1 )$ is the matrix representation of
$ a_t$  in  $ {\cal H}^{2 \times 2}$ defined in {\rm (2.4.1)} and
$$
P_{5,0} = \frac{1}{2} \left[ \begin{array}{cc} (1 + e_{[5]})P_{4,0} &  - (1  - e_{[5]})P_{4,0} \\
 (1 - e_{[5]})P_{4,0} &  (1 + e_{[5]})P_{4,0} \end{array} \right],  \eqno (2.5.4)
 $$
$$
 P_{5,0}^{-1} = \frac{1}{2} \left[ \begin{array}{cc} P_{4,0}^{-1}(1 + e_{[5]}) &  P_{4,0}^{-1}
(1 - e_{[5]}) \\  - P_{4,0}^{-1}(1 - e_{[5]}) &  P_{4,0}^{-1}(1 + e_{[5]}) \end{array} \right],
  \eqno (2.5.5)
$$
$ P_{4,0} $ is given in Theorem 2.4.1. }\\ 

\noindent  {\bf Proof.} \  Note that $ be_{[5]} = e_{[5]}b $ holds for all $ b \in {\cal R}_{4,0}$. By 
applying (2.4.1)  to $  a I_2 =  a_0 I_2 + a_1e_{[5]}I_2$, we get 
\begin{eqnarray*}
P_{4,0}(aI_2) P_{4,0}^{-1} 
& = & P_{4,0}(a_0I_2) P_{4,0}^{-1} +  P_{4,0}(a_1I_2) P_{4,0}^{-1}e_{[5]} = \phi_{4,0}(a_0) + \phi_{4,0}(a_1)e_{[5]}=  \psi(a),  
\end{eqnarray*}
and 
$$ 
P_{4,0}( \overline{a} I_2) P_{4,0}^{-1} = \phi_{4,0}(a_0) - \phi_{4,0}(a_1)e_{[5]}=  \psi(\overline{a}).$$
By Theorem 2.1.1, we construct a matrix and its inverse as follows 
$$ 
V =  \frac{1}{2} \left[ \begin{array}{cc} (1 + e_{[5]})I_2 &  -( 1 - e_{[5]})I_2 \\  (1 - e_{[5]})I_2 &  (1 + e_{[5]}) I_2\end{array} \right], \ \ \ V^{-1} = \frac{1}{2} \left[ \begin{array}{cc} 
(1 + e_{[5]})I_2 &  ( 1 - e_{[5]})I_2 \\  -(1 - e_{[5]})I_2 &  -(1 + e_{[5]})I_2 \end{array} \right].
$$
Now applying them to diag$( \, \psi(a), \  \psi(\overline{a}) \, )$ we obtain
\begin{eqnarray*}
V \left[ \begin{array}{cc} \psi(a)  &  O \\ O &  \psi(\overline{a}) \end{array} \right] V^{-1}
 =\left[ \begin{array}{cc} \phi_{4,0}(a_0) + \phi_{4,0}(a_1) & O  \\ O   & \phi_{4,0}(a_0) - \phi_{4,0}(a_1) \end{array} \right].
\end{eqnarray*}
Finally substituting $ \psi(a) = P_{4,0}(aI_2) P_{4,0}^{-1}$ and  $ \psi(\overline{a}) = 
P_{4,0}(\overline{a}I_2)P_{4,0}^{-1}$ into the left-hand side of the above equality produces Eq.(2.5.3). \qquad  $ \Box $ \\

\noindent {\bf Theorem 2.5.2.} \ {\em Let $ a \in {\cal R}_{4,1} = {\cal R} \{ e_1, \ e_2, \ e_3, \ e_4,\
 \varepsilon_1  \},$ and write $a$ as $a=  a_0 + a_1( e_{[4]} \varepsilon_1), $ where 
$$
a_0, \ a_1 \in {\cal R}_{3,1} = {\cal R}\{ \, e_1, \ e_2, \ e_3,
\ \varepsilon_1 \, \},
\qquad ( e_{[4]} \varepsilon_1 )^2  = - 1.  \
$$
Then $ a I_4$ satisfies the following universal similarity factorization  equality 
$$
 P_{4,1}(aI_4) P_{4,1}^{-1} =  \phi_{3,1}(a_0) +  \phi_{3,1}(a_1)(e_{[4]} \varepsilon_1) 
\equiv \phi_{4,1}(a) \in {\cal C }^{ 4 \times 4},   \eqno (2.5.6)
$$ 
where  $ \phi_{3,1}(a_t)( t = 0, \ 1 )$ is the matrix representation of $ a_t$ in $ {\cal R}^{4 \times 4}$ given in {\rm (2.4.3)} and $  P_{4,1} =  P_{3,1},$  the matrix given in {\rm(2.4.4)}. }\\

\noindent {\bf Proof.} \  Follows directly from Eq.(2.4.3). \qquad  $\Box $ \\

For economizing space,  we omit  the proofs of all the  following several results. \\

\noindent {\bf Theorem 2.5.3.} \ {\em Let $ a \in {\cal R}_{3,2} = {\cal R}\{ e_1, \ e_2, \ e_3,
  \ \varepsilon_1, \ \varepsilon_2 \},$ and write $a$ as $ a=  a_0 + a_1 (e_{[3]}\varepsilon_{[2]}),$
where 
$$
a_0, \ a_1 \in {\cal R}_{2,2} = {\cal R}\{ \, e_1, \ e_2, \ \varepsilon_1, \
\varepsilon_2 \,\},  \qquad (e_{[3]}\varepsilon_{[2]})^2  = 1.  \eqno (2.5.7)
$$
Then the diagonal matrix $ D_a = {\rm diag} ( aI_2 , \  \overline{a}I_2 ) $ satisfies the following universal similarity factorization  equality 
$$
 P_{3,2}D_a P_{3,2}^{-1}
 = \left[ \begin{array}{cc} \phi_{2,2}(a_0) + \phi_{2,2}(a_1) & O  \\ O   & \phi_{2,2}(a_0) - \phi_{2,2}(a_1) \end{array} \right] \equiv \phi_{3,2}(a) \in \, ^2{\cal R }^{ 4 \times 4},   \eqno (2.5.8)
$$ 
where  $ \phi_{2,2}(a_t)( t = 0, \ 1 )$ is the matrix representation of $ a_t$ in $ {\cal R}^{4\times 4}$
 defined in {\rm (2.4.5)} and 
$$
P_{3,2} = \frac{1}{2} \left[ \begin{array}{cc} (1 + e_{[3]}\varepsilon_{[2]})P_{2,2} &
  - (1  - e_{[3]}\varepsilon_{[2]})P_{2,2} \\
 (1 - e_{[3]}\varepsilon_{[2]})P_{2,2} & ( 1 + e_{[3]}\varepsilon_{[2]})P_{2,2} \end{array} \right],  \eqno (2.5.9)
$$
$$
P_{3,2}^{-1} = \frac{1}{2} \left[ \begin{array}{cc} P_{2,2}^{-1} (1 + e_{[3]}\varepsilon_{[2]})
 & P_{2,2}^{-1}(1 - e_{[3]}\varepsilon_{[2]}) \\  - P_{2,2}^{-1}(1 - e_{[3]}\varepsilon_{[2]}) &  P_{2,2}^{-1}( 1 + e_{[3]}\varepsilon_{[2]}) \end{array} \right],  \eqno (2.5.10)
$$
$ P_{2,2} $ and $ P_{2,2}^{-1} $ are given in Eq.{\rm (2.4.6)}. }\\

\noindent {\bf Theorem 2.5.4.} \ {\em Let $ a \in {\cal R}_{2,3} = {\cal R} \{ e_1, \ e_2, \ \varepsilon_1, \ \varepsilon_2, \ \varepsilon_3 \},$ and write $a$ as $ a=  a_0 + a_1 (e_{[2]} \varepsilon_{[3]}),$ 
where 
$$
a_0, \ a_1 \in {\cal R}_{2,2} = {\cal R}\{ \, e_1, \ e_2, \ \varepsilon_1, \ \varepsilon_2 \, \}, \qquad ( e_{[2]} \varepsilon_{[3]} )^2  = - 1.  
$$
Then $ a I_4$ satisfies the following universal similarity equality 
$$
 P_{2,3}(aI_4) P_{2,3}^{-1} =  \phi_{2,2}(a_0) +  \phi_{2,2}(a_1)e_{[2]} \varepsilon_{[3]} 
\equiv \phi_{2,3}(a) \in {\cal C }^{ 4 \times 4},  \eqno (2.5.11)
$$ 
where  $ \phi_{2,2}(a_t)( t = 0, \ 1 )$ is the matrix representation of $ a_t$ in $ {\cal R}^{4 \times 4}$
 defined in Eq.{\rm (2.4.5)} and $  P_{2,3} =  P_{2,2},$  the matrix given in Eq.{\rm (2.4.6)}. }\\

 \noindent {\bf Theorem 2.5.5.} \ {\em Let $ a \in {\cal R}_{1,4} = {\cal R}\{ e_1, \ \varepsilon_1, \ \varepsilon_2,  \ \varepsilon_3, \ \varepsilon_4  \},$ and write $a$ as $a=  a_0 + a_1 (e_1\varepsilon_{[4]}), $ 
where 
$$
a_0, \ a_1 \in {\cal R}_{1,3} = {\cal R}\{ \, e_1, \ \varepsilon_1, \
\varepsilon_2, \ \varepsilon_3 \, \},  \qquad (e_1\varepsilon_{[4]})^2  = 1.
$$
Then the diagonal matrix $ D_a = {\rm diag}( aI_2 , \  \overline{a}I_2 ) $ satisfies the following universal  similarity  factorization equality 
$$
 P_{1,4}D_aP_{1,4}^{-1}
 = \left[ \begin{array}{cc} \phi_{1,3}(a_0) + \phi_{1,3}(a_1) & O  \\ O   & \phi_{1,3}(a_0) - \phi_{1,3}(a_1) \end{array} \right] \equiv \phi_{1,4}(a) \in \, ^2{\cal H }^{ 2 \times 2}, \eqno (2.5.12)
$$ 
where  $ \phi_{1,3}(a_t)( t = 0, \ 1 )$ is the matrix representation of $ a_t$ in $ {\cal H}^{2 \times 2}$
 defined in Eq.{\rm (2.4.7)} and 
$$
P_{1,4} = \frac{1}{2} \left[ \begin{array}{cc} (1 + e_1\varepsilon_{[4]})P_{1,3} &
  - (1  - e_1\varepsilon_{[4]})P_{1,3} \\
 (1 - e_1\varepsilon_{[4]})P_{1,3} & ( 1 + e_1\varepsilon_{[4]})P_{1,3} \end{array} \right],  
\eqno (2.5.13)
$$
$$
P_{1,4}^{-1} = \frac{1}{2} \left[ \begin{array}{cc} P_{1,3}^{-1} (1 + e_1\varepsilon_{[4]})
 & P_{1,3}^{-1}(1 - e_1\varepsilon_{[4]}) \\  - P_{1,3}^{-1}(1 - e_1\varepsilon_{[4]}) &  P_{1,3}^{-1}( 1 + e_1\varepsilon_{[4]}) \end{array} \right], \eqno (2.5.14)
$$
$ P_{1,3} $ and $ P_{1,3}^{-1} $ are given in Eq.{\rm (2.4.8)}. }\\

\noindent {\bf Theorem 2.5.6.} \ { \em Let $ a \in {\cal R}_{0,5} = {\cal R} \{ \varepsilon_1, \ \cdots, \ 
\varepsilon_5 \}.$
Then $a$ can factor as $ a=  a_0 + a_1 \varepsilon_{[5]},$
where 
$$
a_0, \ a_1 \in {\cal R}_{2,2} = {\cal R}\{ \, \varepsilon_{1234}, \
 \varepsilon_{1235}, \ \varepsilon_1, \ \varepsilon_2 \, \}, \qquad \varepsilon_{[5]}^2  = - 1.
$$
Then  there is  an independent invertible matrix $ P_{0,5}$  over $ {\cal R}_{0,5}$ such that
  $ a I_4$ satisfies the following universal similarity equality 
$$
 P_{0,5}(aI_4) P_{0,5}^{-1} =  \phi_{2,2}(a_0) +  \phi_{2,2}(a_1) \varepsilon_{[5]} 
\equiv \phi_{0,5}(a) \in {\cal C }^{ 4 \times 4},  \eqno (2.5.15)
$$ 
where  $ \phi_{2,2}(a_t)( t = 0, \ 1 )$ is the matrix representation of $ a_t$ in $ {\cal R}^{4 \times 4}$
 defined in Eq.{\rm (2.4.5)}. }\\

\noindent {\bf 2.6. The Cases for $ {\cal R}_{p,q}$ with $ p+ q= 6$ } \\

According to the table Eq.(1.4), we know that the seven algebraic isomorphisms for the  Clifford algebras ${\cal R}_{p,q}$ with $ p+ q= 6 $ are as follows 
$$ 
{\cal R}_{6,0} \  \simeq   \ {\cal H}^{4 \times 4}, \ \ \ {\cal R}_{5,1} \  \simeq  \ {\cal H}^{4 \times 4}, \ \ \  
{\cal R}_{4,2} \  \simeq  \ {\cal R}^{8 \times 8}, \ \ \ {\cal R}_{3,3} \ \simeq \ {\cal R}^{8 \times 8},   \eqno (2.6.1)
$$
$$
{\cal R}_{2,4} \  \simeq \ {\cal H}^{4 \times 4},  \qquad
{\cal R}_{1,5} \  \simeq \  {\cal H}^{4 \times 4},  \qquad
{\cal R}_{0,6} \  \simeq  \ {\cal R}^{8 \times 8}.   \eqno (2.6.2)
$$

For economizing space,  we omit  the proofs of all the results in this
subsection. \\

\noindent {\bf Theorem 2.6.1.} \ {\em Let $ a \in {\cal R}_{6,0} = {\cal R}\{ \, e_1, \ \cdots, \ e_6 \,
 \}.$ Then $a$ can factor as
$$
a  =  a_0 + a_1(e_{[4]}e_5) + a_2(e_{[4]}e_6) + a_3e_{56} = a_0 + (e_{[4]}e_5)a_1 + (e_{[4]}e_6)a_2 + e_{56}a_3,  
$$
where 
$$
a_0, \ a_1, \ a_2, \ a_3 \in {\cal R}_{4,0} = {\cal R}\{ \, e_1, \ e_2,  \ e_3, \ e_4 \, \}
 $$
$$ 
(e_{[4]}e_5)^2 = 1, \qquad   (e_{[4]}e_6)^2 = 1, \qquad
 e_{56}= (e_{[4]}e_5 )(e_{[4]}e_6)=  - (e_{[4]}e_6)(e_{[4]}e_5 ).
$$
Moreover$,$ there is an independent invertible matrix $ P_{6,0}$  over  $ {\cal R}_{6,0}$ such that
  $ a I_4$ satisfies the following universal similarity  factorization equality 
$$
 P_{6,0}( aI_4)P_{6,0}^{-1}  = \left[ \begin{array}{cc}  \phi_{4,0}(a_0) + \phi_{4,0}(a_1)  & 
 \phi_{4,0}(a_2) + \phi_{4,0}(a_3)
 \\ \phi_{4,0}(a_2) - \phi_{4,0}(a_3)   & \phi_{4,0}(a_0) - \phi_{4,0}(a_1) \end{array} \right] 
\equiv \phi_{6,0}(a) \in {\cal H}^{4 \times 4},
$$
where $ \phi_{4,0}(a_t)( t = 0$---$3)$ is the matrix representation of $ a_t$  in  $ {\cal H}^{2 \times 2}$ 
 defined  in {\rm (2.4.1)}.  } \\ 

\noindent {\bf Theorem 2.6.2.} \ {\em Let $ a \in {\cal R}_{5,1} = {\cal R}\{ e_1, \ \cdots, \  e_5,
 \ \varepsilon_1 \}.$ Then $a$ can factor as
$$
a  =  a_0 + a_1 e_{[5]} + a_2(e_{[4]} \varepsilon_1) + a_3 (e_5 \varepsilon_1)  = a_0 + e_{[5]}a_1 + (e_{[4]} \varepsilon_1)a_2 + (e_5 \varepsilon_1)a_3,  
$$
where 
$$
a_0, \ a_1, \ a_2, \ a_3 \in {\cal R}_{4,0} =
{\cal R}\{ \, e_1, \ e_2,  \ e_3, \ e_4 \,\},
$$
$$ 
e_{[5]}^2 = 1, \qquad  (e_{[4]} \varepsilon_1)^2 = -1, \qquad 
 e_5 \varepsilon_1 = e_{[5]}(e_{[4]} \varepsilon_1)
 = -(e_{[4]} \varepsilon_1)e_{[5]}.
$$
Moreover$,$ there is an independent invertible matrix $ P_{5,1}$ over $ {\cal R}_{5,1}$ such that
  $ a I_4$ satisfies the following  universal similarity  factorization equality 
$$
 P_{5,1}( aI_4)P_{5,1}^{-1}  = \left[ \begin{array}{cc}  \phi_{4,0}(a_0) + \phi_{4,0}(a_1)  & 
 -[\, \phi_{4,0}(a_2) + \phi_{4,0}(a_3)\,]  \\ \phi_{4,0}(a_2) - \phi_{4,0}(a_3)   & \phi_{4,0}(a_0) - \phi_{4,0}(a_1) \end{array} \right] \equiv \phi_{5,1}(a) \in {\cal H}^{4 \times 4},
$$
where $ \phi_{4,0}(a_t)( t = 0$---$3)$ is the matrix representation of $ a_t$  in  $ {\cal H}^{2 \times 2}$ 
 defined  in Eq.{\rm (2.4.1)}. } \\

\noindent {\bf Theorem 2.6.3.} \ {\em Let $ a \in {\cal R}_{4,2} = {\cal R}\{ e_1, \ \cdots, \ \ e_4, \ 
\varepsilon_1, \ \varepsilon_2 \}.$   
$$
a  = a_0 + a_1 (e_{[3]} \varepsilon_{[2]}) + a_2(e_{[4]} \varepsilon_1) +
 a_3 (e_4 \varepsilon_2) = a_0 +  (e_{[3]} \varepsilon_{[2]})a_1 + (e_{[4]} \varepsilon_1)a_2 +
  (e_4 \varepsilon_2)a_3,  
$$
where 
$$
a_0, \ a_1, \ a_2, \ a_3 \in {\cal R}_{3,1}
= {\cal R}\{  \, e_1, \ e_2,  \ e_3, \ \varepsilon_1 \,\},
$$
$$ 
(e_{[3]} \varepsilon_{[2]}) ^2 = 1, \ \ \ \ (e_{[4]} \varepsilon_1)^2 = -1, \qquad
 e_4 \varepsilon_2 = ( e_{[3]}\varepsilon_{[2]}) (e_{[4]}\varepsilon_1)
= -(e_{[4]} \varepsilon_1)( e_{[3]}\varepsilon_{[2]}).
$$
Morever$,$ there is an independent invertible matrix $ P_{4,2}$ over $ {\cal R}_{4,2}$ such that
  $ a I_8$ satisfies the following  universal similarity  factorization equality 
$$
 P_{4,2}( aI_8)P_{4,2}^{-1}  = \left[ \begin{array}{cc}  \phi_{3,1}(a_0) + \phi_{3,1}(a_1)  & 
 -[\, \phi_{3,1}(a_2) + \phi_{3,1}(a_3)\,]  \\ \phi_{3,1}(a_2) - \phi_{3,1}(a_3)   & \phi_{3,1}(a_0) - 
\phi_{3,1}(a_1) \end{array} \right] \equiv \phi_{4,2}(a) \in {\cal R}^{8 \times 8},
$$
where $ \phi_{3,1}(a_t)( t = 0$---$3)$ is the matrix representation of $ a_t$  in  $ {\cal R}^{4 \times 4}$ 
 defined  in Eq.{\rm (2.4.3)}. } \\

\noindent {\bf Theorem 2.6.4.} \ {\em Let $ a \in {\cal R}_{3,3} = {\cal R}\{ e_1, \ e_2,  \ e_3, \ 
\varepsilon_1, \ \varepsilon_2, \ \varepsilon_3 \}.$ Then  $a$ cn factor as    
$$
a = a_0 + a_1 (e_{[3]} \varepsilon_{[2]}) + a_2(e_{[2]} \varepsilon_{[3]}) +  a_3 (e_3 \varepsilon_3)  =  a_0 +(e_{[3]} \varepsilon_{[2]})a_1+ (e_{[2]} \varepsilon_{[3]})a_2
 + (e_3 \varepsilon_3)a_3,
$$
where 
$$
a_0, \ a_1, \ a_2, \ a_3 \in {\cal R}_{2,2} = {\cal R}\{ \, e_1, \ e_2,
\ \varepsilon_1, \ \varepsilon_2 \, \},
$$
$$ 
(e_{[3]} \varepsilon_{[2]})^2 = 1, \qquad (e_{[2]} \varepsilon_{[3]})^2 = -1, \qquad  e_3 \varepsilon_3 = ( e_{[3]}\varepsilon_{[2]})(e_{[2]} \varepsilon_{[3]}) 
= -(e_{[2]} \varepsilon_{[3]})( e_{[3]}\varepsilon_{[2]}).
$$
Moreover there is an independent invertible matrix $ P_{3,3}$ over $ {\cal R}_{3,3}$ such that
  $ a I_8$ satisfies the following universal similarity factorization  equality 
$$
 P_{3,3}( aI_8)P_{3,3}^{-1}  = \left[ \begin{array}{cc}  \phi_{2,2}(a_0) + \phi_{2,2}(a_1)  & 
 -[ \phi_{2,2}(a_2) + \phi_{2,2}(a_3)]  \\ \phi_{2,2}(a_2) - \phi_{2,2}(a_3)   & \phi_{2,2}(a_0) - 
\phi_{2,2}(a_1) \end{array} \right] \equiv \phi_{3,3}(a) \in {\cal R}^{8 \times 8},
$$
where $ \phi_{2,2}(a_t)( t = 0$---$3)$ is the matrix representation of $ a_t$ in $ {\cal R}^{4 \times 4}$ 
 defined  in {\rm (2.4.5)}. } \\

\noindent {\bf Theorem 2.6.5.} \ {\em Let $ a \in {\cal R}_{2,4} = {\cal R}\{ e_1, \ e_2,  \  \ 
\varepsilon_1, \ \varepsilon_2, \ \varepsilon_3, \  \varepsilon_4  \}.$ Then
 $a$ can factor as 
$$
a = a_0 + a_1 (e_1 \varepsilon_{[4]}) + a_2(e_{[2]} \varepsilon_{[3]})
 +  a_3 (e_4 \varepsilon_2)  =  a_0 + (e_1 \varepsilon_{[4]})a_1 + (e_{[2]} \varepsilon_{[3]})a_2 +  (e_4 \varepsilon_2)a_3,  
$$
where 
$$
a_0, \ a_1, \ a_2, \ a_3 \in {\cal R}_{1,3}
= {\cal R}\{ \, e_1, \ \varepsilon_1, \ \varepsilon_2, \varepsilon_3 \, \},
$$
$$ 
(e_1 \varepsilon_{[4]})^2 = 1, \qquad (e_{[2]} \varepsilon_{[3]})^2 = -1, \qquad  e_4 \varepsilon_2 = ( e_1\varepsilon_{[4]})(e_{[2]} \varepsilon_{[3]})
 = -(e_{[2]} \varepsilon_{[3]})( e_1\varepsilon_{[4]}).
$$
Moreover$,$ there is an independent invertible matrix $ P_{2,4}$ over $ {\cal R}_{2,4}$ such that
  $ a I_4$ satisfies the following  universal similarity  factorization equality 
$$
 P_{2,4}( aI_4)P_{2,4}^{-1}  = \left[ \begin{array}{cc}  \phi_{1,3}(a_0) + \phi_{1,3}(a_1)  & 
 -[ \, \phi_{1,3}(a_2) + \phi_{1,3}(a_3) \, ]  \\ \phi_{1,3}(a_2) - \phi_{1,3}(a_3)   & \phi_{1,3}(a_0) - 
\phi_{1,3}(a_1) \end{array} \right] \equiv \phi_{2,4}(a) \in {\cal H}^{4 \times 4},
$$
where $ \phi_{1,3}(a_t)( t = 0$---$3)$ is the matrix representation of $ a_t$ in $ {\cal H}^{2 \times 2}$ 
 defined  in Eq.{\rm (2.4.7)}. } \\

\noindent {\bf Theorem 2.6.6.} \ {\em Let $ a \in {\cal R}_{1,5}
= {\cal R}\{ e_1, \ \varepsilon_1, \ \cdots, \  \varepsilon_5 \}.$ Then
 $a$ can factor as
$$
a = a_0 + a_1 (e_1 \varepsilon_{[4]}) + a_2 \varepsilon_{[5]} + a_3(e_1 \varepsilon_5)  = a_0 + (e_1 \varepsilon_{[4]})a_1 + \varepsilon_{[5]}a_2
 + (e_1 \varepsilon_5)a_3, 
$$
where 
$$
a_0, \ a_1, \ a_2, \ a_3 \in {\cal R}_{0,4}
= {\cal R}\{ \, \varepsilon_1, \ \varepsilon_2, \varepsilon_3, \
\varepsilon_4 \, \}, 
$$
$$ 
(e_1 \varepsilon_{[4]})^2 = 1, \qquad \varepsilon_{[5]}^2 = -1, \qquad 
 e_1 \varepsilon_5 = ( e_1\varepsilon_{[4]})\varepsilon_{[5]} = -\varepsilon_{[5]}
( e_1\varepsilon_{[4]}).
$$
Moreover$,$ there is an independent invertible matrix $ P_{1,5}$ over $ {\cal R}_{1,5}$ such that
  $ a I_4$ satisfies the following universal  similarity factorization  equality 
$$
 P_{1,5}( aI_4)P_{1,5}^{-1}  = \left[ \begin{array}{cc}  \phi_{0,4}(a_0) + \phi_{0,4}(a_1)  & 
 -[ \phi_{0,4}(a_2) + \phi_{0,4}(a_3)]  \\ \phi_{0,4}(a_2) - \phi_{0,6}(a_3)   & \phi_{0,4}(a_0) - 
\phi_{0,4}(a_1) \end{array} \right] \equiv \phi_{1,5}(a) \in {\cal H}^{4 \times 4},
$$
where $ \phi_{0,4}(a_t)( t = 0$---$3)$ is the matrix representation of $ a_t$ in $ {\cal H}^{2 \times 2}$ 
 defined  in Eq.{\rm (2.4.9)}. } \\ 

\noindent {\bf Theorem 2.6.7.} \ {\em Let $ a \in {\cal R}_{0,6} =
 {\cal R}\{  \varepsilon_1, \ \cdots, \ \varepsilon_6  \}.$ Then $a$ can
 factor  as 
$$
a = a_0 + a_1 (\varepsilon_{[3]}\varepsilon_6) + a_2(\varepsilon_{[3]}\varepsilon_5) + 
 a_3\varepsilon_{56}  = a_0 + (\varepsilon_{[3]}\varepsilon_6) a_1 + (\varepsilon_{[3]}\varepsilon_5)a_2  + \varepsilon_{56}a_3,
$$
where 
$$
a_0, \ a_1, \ a_2, \ a_3 \in {\cal R}_{3,1} = {\cal R}\{ \, \varepsilon_{124}, \ \varepsilon_{134}, \  \varepsilon_{234}, \ 
\varepsilon_{[3]}\varepsilon_{56} \, \}, 
$$
$$ 
 (\varepsilon_{[3]}\varepsilon_6)^2 = 1, \ \ \ (\varepsilon_{[3]}\varepsilon_5)^2 = 1, 
\ \ \ \varepsilon_{56} = (\varepsilon_{[3]}\varepsilon_6)(\varepsilon_{[3]}\varepsilon_5) = -(\varepsilon_{[3]}\varepsilon_5)(\varepsilon_{[3]}\varepsilon_6).
$$
Moreover$,$ there is an independent invertible matrix $ P_{0,6}$ over $ {\cal R}_{0,6}$ such that
  $ a I_8$ satisfies the following universal  similarity equality 
$$
 P_{0,6}( aI_8)P_{0,6}^{-1}  = \left[ \begin{array}{cc}  \phi_{3,1}(a_0) + \phi_{3,1}(a_1)  & 
 \phi_{3,1}(a_2) + \phi_{3,1}(a_3) \\ \phi_{3,1}(a_2) - \phi_{3,1}(a_3)   & \phi_{3,1}(a_0) - 
\phi_{3,1}(a_1) \end{array} \right] \equiv \phi_{0,6}(a) \in {\cal R}^{8 \times 8},
$$
where $ \phi_{3,1}(a_t)( t = 0$---$3)$ is the matrix representation of $ a_t$ in $ {\cal R}^{4 \times 4}$ 
 defined  in Eq.{\rm (2.4.3)}. } \\ 

\noindent {\bf 2.7. The Cases for $ {\cal R}_{p,q}$ with  $ p +q = 7$ } \\

According to the table (1.4),  the eight algebraic isomorphisms for the  Clifford algebras ${\cal R}_{p,q}$ with $ p+ q= 7 $ are as follows 
$$ 
{\cal R}_{7,0} \  \simeq   \ {\cal C}^{8 \times 8}, \ \ \
{\cal R}_{6,1} \  \simeq  \ ^2{\cal H}^{4 \times 4}, \ \ \  
{\cal R}_{5,2} \  \simeq  \ {\cal C}^{8 \times 8}, \ \ \ 
{\cal R}_{4,3} \ \simeq \ ^2{\cal R}^{8 \times 8},   \eqno (2.7.1)
$$
$$
{\cal R}_{3,4} \  \simeq \ {\cal C}^{8 \times 8},   \ \ \
{\cal R}_{2,5} \  \simeq \  ^2{\cal H}^{4 \times 4},  \ \ \  
{\cal R}_{1,6} \  \simeq  \ {\cal C}^{8 \times 8}, \ \ \
{\cal R}_{0,7} \  \simeq  \ ^2{\cal R}^{8 \times 8}.   \eqno (2.7.2)
$$
The derivation of the universal similarity equalities between elements of ${\cal R}_{p,q}$ with  $ p+ q= 7 $  and the eight matrix algebras are much analogous to those in Subsection 2.5. Here we only present the results for ${\cal R}_{7,0}$ and 
 ${\cal R}_{0,7}$. \\

Just as for ${\cal R}_{p,q}$ with $ p+ q= 5$, we can decompose $ a \in {\cal R}_{p,q}$
 with $ p+ q= 7 $ into the following general form
$$ 
a = a_0 + a_1e_{[7]} = a_0 + e_{[7]}a_1,  \eqno (2.7.3)
$$ 
where $ a_0, \ a_1$ are elements of the Clifford algebras defined on any six generators of $ e_1, \  e_2,
\ \cdots, \  e_7$.  and from this decomposition, we introduce the conjugate of $ a $ as follows
$$
 \overline{a} =  a_0 - a_1e_{[7]}. \eqno (2.7.4)
$$
Then it is easy to verify that for all $ a, \ b \in {\cal R}_{p,q},$ and
 $ p + q = 7,  \ \lambda \in {\cal R}$,
$$ 
\overline{\overline{a}} = a,  \ \ \  \overline{ a + b } = \overline{ a } + \overline{ b }, \ \ \ 
\overline{ab}=\overline{a}\overline{b}, \ \ \  \overline {\lambda a} = \overline { a \lambda } = 
 \lambda \overline{a}. \eqno (2.7.5) 
$$

\noindent {\bf Theorem 2.7.1.} \ {\em Let $ a \in {\cal R}_{7,0} = {\cal R}\{ e_1, \ \cdots, \ e_7 \},$ and write $a$ as $  a=  a_0 + a_1e_{[7]},$ where $ e_{[7]}^2  = -1$, and 
$$
a_0, \ a_1 \in {\cal R}_{4,2} = {\cal R} \{ \,  e_1, \ e_2, \ e_3,  \ e_4,
\ e_{[5]}e_6, \ e_{[5]}e_7 \, \}. 
$$
Then there  is an independent invertible matrix $ P_{7,0}$ over $ {\cal R}_{7,0}$ such that
  $ aI_8$ satisfies the following universal similarity  factorization equality 
$$
 P_{7,0}(aI_8)P_{7,0}^{-1} = \phi_{4,2}(a_0) +  \phi_{4,2}(a_1)e_{[7]}  \equiv \phi_{7,0}(a) \in {\cal C }^{ 8 \times 8},
$$ 
where  $ \phi_{4,2}(a_t)( t = 0, \ 1 )$ is the matrix representation of $ a_t$  in  $ {\cal R}^{8 \times 8}$  defined in Theorem 2.6.3. } \\

\noindent {\bf Theorem 2.7.2.} \ {\em Let $ a \in {\cal R}_{0,7} = {\cal R} \{ \varepsilon_1, \ \cdots, \ 
\varepsilon_7  \},$ and write $a$ as $ a=  a_0 + a_1 \varepsilon_{[7]},$  where  $ \varepsilon_{[7]}^2  = 1$, and 
$$
a_0, \ a_1 \in {\cal R}_{0,6} = {\cal R}\{ \, \varepsilon_1, \ \cdots,
\ \varepsilon_6 \,\}.
$$
Then  there is an independent invertible matrix $ P_{0,7}$ over $ {\cal R}_{0,7}$ such that the diagonal matrix $ D_a = {\rm diag} ( aI_8, \  \overline{a}I_8 ) $ satisfies the following similarity equality 
$$
 P_{0,7}D_a P_{0,7}^{-1}
 = \left[ \begin{array}{cc} \phi_{0,6}(a_0) + \phi_{0,6}(a_1) & O  \\ O   & \phi_{0,6}(a_0) - \phi_{0,6}(a_1) \end{array} \right] \equiv \phi_{0,7}(a) \in \, ^2{\cal R }^{ 8 \times 8}, 
$$ 
where  $ \phi_{0,6}(a_t)( t = 0, \ 1 )$ is the matrix representation of $ a_t$  in  $ {\cal R}^{8 \times 8}$ defined in Theorem 2.6.7. } \\

\noindent {\bf 2.8. The Cases for $ {\cal R}_{p,q}$ with $ p +q = 8$ } \\

Just as in subsection 2.7, we only examine the two particular  cases over ${\cal R}_{8,0}$ and 
 $ {\cal R}_{0,8}$. \\

\noindent {\bf Theorem 2.8.1.} \ {\em Let $ a \in {\cal R}_{8,0} =
 {\cal R}\{ e_1, \ \cdots, \ e_8 \}.$ Then $a$ can factor as 
$$
a = a_0 + a_1e_{4567} + a_2e_{4568} + a_3e_{78}  = a_0 + e_{4567}a_1 + e_{4568} a_2+ e_{78}a_3,   
$$
where 
$$
a_t  \in {\cal R}_{3,3} = {\cal R}\{ e_1, \ e_2,  \ e_3, \ e_{[3]}e_{478}, \  
e_{[3]}e_{578}, \ e_{[3]}e_{678} \ | \ 1, \ 1, \ 1, \  -1 \,  -1, \ -1 \}, 
$$
$$ 
e_{4567}^2 = 1, \qquad  e_{4568}^2 = 1, \qquad  e_{78} = (e_{4567})(e_{4568})= -( e_{4568})(e_{4567}),
$$ 
$ t =0$---$3$. In that case$,$ there is an independent invertible matrix $ P_{8,0}$  over  $ {\cal R}_{8,0}$ such that
  $ a I_{16}$ satisfies the following universal similarity factorization  equality 
$$
 P_{8,0}( aI_{16})P_{8,0}^{-1}  = \left[ \begin{array}{cc}  \phi_{3,3}(a_0) + \phi_{3,3}(a_1)  & 
 \phi_{3,3}(a_2) + \phi_{3,3}(a_3)
 \\ \phi_{3,3}(a_2) - \phi_{3,3}(a_3)   & \phi_{3,3}(a_0) - \phi_{3,3}(a_1) \end{array} \right] 
\equiv \phi_{8,0}(a) \in {\cal R}^{16 \times 16},
$$
where $ \phi_{3,3}(a_t)( t = 0$---$3)$ is the matrix representation of $ a_t$  in  $ {\cal R}^{8 \times 8}$ 
 defined  in Theorem 2.6.4.} \\ 

\noindent {\bf Theorem 2.8.2.} \ {\em Let $ a \in {\cal R}_{0,8} = {\cal R} \{ \varepsilon_1, \ \cdots, \ 
\varepsilon_8 \},$ Then $a$ can factor as   
$$
a  = a_0 + a_1 (\varepsilon_{[6]}\varepsilon_8) + a_2 ( \varepsilon_{[6]}\varepsilon_7) +
 a_3\varepsilon_{78}  = a_0 + (\varepsilon_{[6]}\varepsilon_8)a_1 +  (\varepsilon_{[6]}\varepsilon_7) a_2 +
 \varepsilon_{78}a_3 
$$
where 
$$
a_0, \ a_1, \ a_2, \  a_3 \in {\cal R}_{0,6} = {\cal R}\{ \, \varepsilon_1,
\ \cdots,  \ \varepsilon_6 \, \}, 
$$
and 
$$
(\varepsilon_{[6]}\varepsilon_8)^2 = 1, \qquad  ( \varepsilon_{[6]}\varepsilon_7)^2 = 1, 
\qquad \varepsilon_{78} = (\varepsilon_{[6]}\varepsilon_8) ( \varepsilon_{[6]}\varepsilon_7) = -(\varepsilon_{[6]}\varepsilon_7) ( \varepsilon_{[6]}\varepsilon_8). 
$$
In that case$,$ there is an independent invertible matrix $ P_{0,8}$ over $ {\cal R}_{0,8}$ such that $ aI_{16}$  satisfies the following  universal similarity factorization  equality 
$$
 P_{0,8}(aI_{16}) P_{0,8}^{-1}
 = \left[ \begin{array}{cc} \phi_{0,6}(a_0) + \phi_{0,6}(a_1) &  \phi_{0,6}(a_2) + \phi_{0,6}(a_3) 
\\ \phi_{0,6}(a_2) - \phi_{0,6}(a_3)    & \phi_{0,6}(a_0) - \phi_{0,6}(a_1) \end{array} \right] \equiv \phi_{0,8}(a) \in {\cal R }^{ 16 \times 16}, 
$$ 
where  $ \phi_{0,6}(a_t)(t = 0$---$3)$ is the matrix representation of $ a_t$  in  $ {\cal R}^{8 \times 8}$ defined in Theorem 2.6.7. } \\

\noindent {\bf 3.  UNIVERSAL SIMILARITY EQUALITIES OVER $ {\cal R}_{n+p,n} $ AND $ {\cal R}_{n,n+q} $ WITH $ 0 \leq p,q \leq 6$ }   \\

On the basis of the results in Section 2, we present in this section several
induction formulas for the universal similarity
factorization equalities over $ {\cal R}_{n+p,n} $ and  $ {\cal R}_{n,n+q}$ with $ 0 \leq p,q \leq 6$. \\

\noindent {\bf 3.1. The Cases for $ {\cal R}_{n,n}$ } \\

We have shown in Theorems 2.2.2, 2.4.3 and 2.6.4  that
 for all $ a \in {\cal R}_{n,n}$ with $ n = 1, \ 2, \ 3$, the
 corresponding $ aI_2, \  aI_4, \ aI_8 $ are uniformly similar to
 their real matrix representations, respectively. By induction,
 we can establish the following general result for the split Clifford algebra
 $ {\cal R}_{n,n}$. \\

\noindent {\bf Theorem 3.1.1.} \ {\em Suppose that there is an independent invertible matrix $ P_{n-1,n-1}
(n \geq 1)$ over  $  {\cal R }_{n-1,n-1} ={ \cal R }\{ \,  e_1, \ \cdots, \ e_{n-1}, \  \varepsilon_1, \ \cdots, \ \varepsilon_{n-1} \, \}$ such that 
$$
 P_{n-1,n-1}( aI_{2^{n-1}}) P_{n-1,n-1}^{-1} \equiv \phi_{n-1,n-1}(a) \in
 {\cal R }^{2^{n-1} \times 2^{n-1}}  \eqno (3.1.1)
$$  
holds for all $ a \in {\cal R}_{n-1,n-1}$. Now let
$ a \in {\cal R}_{n,n}={ \cal R }\{ \,  e_1, \ \cdots, \ e_n, \
\varepsilon_1, \ \cdots, \ \varepsilon_{n} \, \}.$ Then it can factor as
\begin{eqnarray*}
 a &=&  a_0 + a_1(e_{[n]} \varepsilon_{[n-1]}) + a_2(e_{[n-1]} \varepsilon_{[n]}) + a_3\mu_{n,n} \\
 &= &a_0 + (e_{[n]} \varepsilon_{[n-1]})a_1 + (e_{[n-1]} \varepsilon_{[n]})a_2 + \mu_{n,n}a_3,
\end{eqnarray*}
where 
$$
a_0, \ a_1, \ a_2, \  a_3 \in {\cal R }_{n-1,n-1} ={ \cal R }\{ \,  e_1, \ \cdots, \ e_{n-1}, \  \varepsilon_1, \ \cdots, \ \varepsilon_{n-1} \, \},
$$
$$
(e_{[n]} \varepsilon_{[n-1]})^2 = 1, \qquad (e_{[n-1]} \varepsilon_{[n]})^2 =-1, 
$$ 
$$ 
\mu_{n,n} = (e_{[n]} \varepsilon_{[n-1]})(e_{[n-1]} \varepsilon_{[n]}) = -(e_{[n-1]} \varepsilon_{[n]})(e_{[n]} \varepsilon_{[n-1]}) = (-1)^{n-1}e_n\varepsilon_n. 
$$ 
In that case$,$ $ aI_{2^{n}}$ satisfies the following universal similarity
factorization  equality
$$
\displaylines{ 
\hspace*{1cm} 
P_{n,n}( aI_{2^{n}}) P_{n,n}^{-1}  \hfill 
\cr
\hspace*{2cm} 
 = \left[ \begin{array}{cc}  \phi_{n-1,n-1}(a_0) + \phi_{n-1,n-1}(a_1)  &  -[ \, \phi_{n-1,n-1}(a_2) + \phi_{n-1,n-1}(a_3) \, ]  \\ \phi_{n-1,n-1}(a_2) - \phi_{n-1,n-1}(a_3)   & \phi_{n-1,n-1}(a_0) - \phi_{n-1,n-1}(a_1) \end{array} \right]  \hfill 
\cr 
\hspace*{2cm} \equiv \phi_{n,n}(a) \in {\cal R }^{2^n \times 2^n}, \hfill    (3.1.2) 
\cr }
$$
where
$$
P_{n,n}  = \frac{1}{2} \left[ \begin{array}{cc} ( 1 + e_{[n]}\varepsilon_{[n-1]} )P_{n-1,n-1}
 &  ( e_{[n-1]}\varepsilon_{[n]}   - \mu_{n,n} )P_{n-1,n-1} \\ - ( e_{[n-1]}\varepsilon_{[n]} + \mu_{n,n} ) P_{n-1,n-1}  & ( 1 - e_{[n]}\varepsilon_{[n-1]} )P_{n-1,n-1} \end{array} \right],  \eqno (3.1.3)
$$ 
$$
P_{n,n}^{-1}  = \frac{1}{2} \left[ \begin{array}{cc} P_{n-1,n-1}^{-1}( 1 + e_{[n]}\varepsilon_{[n-1]} )
 &  P_{n-1,n-1}^{-1} ( e_{[n-1]}\varepsilon_{[n]}   - \mu_{n,n} ) \\ -  P_{n-1,n-1}^{-1}( e_{[n-1]}\varepsilon_{[n]} + \mu_{n,n} )  &   P_{n-1,n-1}^{-1}( 1 - e_{[n]}\varepsilon_{[n-1]} ) \end{array} \right].  \eqno (3.1.4)
$$ }

\noindent {\bf Proof.} \ Applying Eq.(3.1.1) to $ aI_{2^{n-1}}$,  we obtain 
 \begin{eqnarray*}
 P_{n-1,n-1}( aI_{2^{n-1}}) P_{n-1,n-1}^{-1} &= & \phi_{n-1,n-1}(a_0) +
  \phi_{n-1,n-1}(a_1)(e_{[n]} \varepsilon_{[n-1]}) \\
&  &  + \  \phi_{n-1,n-1}(a_2)(e_{[n-1]} \varepsilon_{[n]}) + \phi_{n-1,n-1}(a_3)\mu_{n,n}  \equiv 
 \psi(a). 
\end{eqnarray*}
Next setting 
$$ 
V  = V^{-1} =  \frac{1}{2} \left[ \begin{array}{cc} ( 1 + e_{[n]}\varepsilon_{[n-1]} )I_{2^{n-1}}
 &  ( e_{[n-1]}\varepsilon_{[n]} - \mu_{n,n} )I_{2^{n-1}} \\ - ( e_{[n-1]}\varepsilon_{[n]} + \mu_{n,n} )I_{2^{n-1}}  & ( 1 - e_{[n]}\varepsilon_{[n-1]} )I_{2^{n-1}} \end{array} \right]. 
$$ 
and applying it to $ D_a ={\rm diag}( \psi(a), \ \psi(a) )$,  we find
$$ 
VD_aV^{-1} = \left[ \begin{array}{cc}  \phi_{n-1,n-1}(a_0) + \phi_{n-1,n-1}(a_1)  &  -[ \, \phi_{n-1,n-1}(a_2) + \phi_{n-1,n-1}(a_3) \, ]  \\ \phi_{n-1,n-1}(a_2) - \phi_{n-1,n-1}(a_3)   & \phi_{n-1,n-1}(a_0) - \phi_{n-1,n-1}(a_1) \end{array} \right].
$$
Finally substituting $ \psi(a) = P_{n-1,n-1}( aI_{2^{n-1}}) P_{n-1,n-1}^{-1}$ into its left-hand side 
yields the desired result.  \qquad  $ \Box $ \\
 
\noindent {\bf 3.2. The Cases for $ {\cal R}_{n+1,n}$ } \\

For $ {\cal R}_{n+1,n}$ with $ n = 0, \ 1, \ 2,$ we have the corresponding universal similarity factorization  equalities in Theorems 2.1.1, 2.2.2 and 2.3.3. In general, we have the following. \\

\noindent {\bf Theorem 3.2.1.} \ {\em Let  $ a \in{ \cal R}_{n+1,n}
={\cal R }\{ \,  e_1, \ \cdots, \ e_{n+1}, \  \varepsilon_1, \ \cdots, \
\varepsilon_{n} \, \} $ be given. Then  $ a $ can factor as 
$$
a = a_0 + a_1e =  a_0 + ea_1,
$$
where 
$$
a_0, \ a_1 \in {\cal R}_{n,n} = {\cal R}\{ \, e_1, \ \cdots,\  e_n, \ \varepsilon_1, \  \cdots, \ \varepsilon_n \, \},
$$
$$
e = e_{[n+1]} \varepsilon_{[n]}, \qquad e^2 = 1. 
$$ 
Moreover define $ \overline{a } = a_0 - a_1e $. In that case$,$ $ D_a = {\rm diag}( \, aI_{2^n}, \ 
\overline{a}I_{2^n} )$  satisfies the following universal similarity factorization  equality 
$$
 P_{n+1,n}D_a P_{n+1,n}^{-1}
 = \left[ \begin{array}{cc} \phi_{n,n}(a_0) + \phi_{n,n}(a_1) & O  \\ O   & \phi_{n,n}(a_0) - \phi_{n,n}(a_1) \end{array} \right] \equiv \phi_{n+1,n}(a) \in \, ^2{\cal R }^{ 2^n \times 2^n},  
$$ 
where  $ \phi_{n,n}(a_t)( t = 0, \ 1 )$ is the matrix representation of $ a_t$ in $ {\cal R}^{2^n \times 2^n }$ defined in Eq.{\rm (3.1.3)} and 
$$
P_{n+1,n} = \frac{1}{2} \left[ \begin{array}{cc} (1 + e)P_{n,n} &
  - (1  - e)P_{n,n} \\ (1 - e)P_{n,n} & ( 1 + e)P_{n,n} \end{array} \right],  \ \ \
P_{n+1,n}^{-1} = \frac{1}{2} \left[ \begin{array}{cc} P_{n,n}^{-1} (1 + e)
 & P_{n,n}^{-1}(1 - e) \\  - P_{n,n}^{-1}(1 - e) &  P_{n,n}^{-1}( 1 + e) \end{array} \right],
 $$
$ P_{n,n} $ and $ P_{n,n}^{-1} $ are the matrices  in Eqs.{\rm (3.1.4)} and {\rm (3.1.5)}. }

\medskip
\noindent {\bf Proof.} \ Applying Eq.(3.1.3) to $  aI_{2^n} =  a_0I_{2^n} + a_1eI_{2^n} $ gives
 \begin{eqnarray*}
 P_{n,n}( aI_{2^n}) P_{n,n}^{-1} & = &   P_{n,n}( aI_{2^n}) P_{n,n}^{-1} +
  P_{n,n}( a_1I_{2^n})P_{n,n}^{-1}e  \\
& = &  \phi_{n,n}(a_0) +  \phi_{n,n}(a_1)e  \equiv \psi(a). 
\end{eqnarray*}
Next setting 
$$ 
V  =  \frac{1}{2} \left[ \begin{array}{cc} ( 1 + e)I_{2^{n}}
 &  - (1 - e) I_{2^{n}} \\   (1 - e) I_{2^{n}} &  ( 1  + e )I_{2^{n}}
 \end{array} \right], \ \ \ \ V^{-1} =  \frac{1}{2} \left[ \begin{array}{cc} ( 1 + e)I_{2^{n}}
 &  (1 - e) I_{2^{n}} \\   -(1 - e) I_{2^{n}} &  ( 1  + e )I_{2^{n}} \end{array} \right],
$$ 
and applying them to $ D_a = {\rm diag}( \, \psi(a), \ \psi(\overline{a}) \, )$, we get 
$$ 
V \left[ \begin{array}{cc}  \psi(a) &  O  \\   O & \psi(\overline{a})  \end{array} \right] V^{-1}=
\left[ \begin{array}{cc} \phi_{n,n}(a_0) + \phi_{n,n}(a_1) & O  \\ O   & \phi_{n,n}(a_0) - \phi_{n,n}(a_1)
 \end{array} \right].
$$
Finally substituting $ \psi(a) = P_{n,n}( aI_{2^{n}}) P_{n,n}^{-1}$ and  $ \psi(\overline{a}) = 
P_{n,n}( \overline{a}I_{2^{n}})P_{n,n}^{-1}$ into its left-hand side 
yields the desired result.  \qquad  $ \Box $ \\

\noindent {\bf 3.3. The Cases for $ {\cal R}_{n+2,n}$ } \\

For $ {\cal R}_{n+2,n}$ with $ n = 0, \ 1, \ 2,$ we have the corresponding universal similarity equalities 
in Theorems 2.2.1, 2.4.2 and 2.6.3. In general, we have the following. 

\medskip

\noindent {\bf Theorem 3.3.1.} \ {\em Suppose that there is an independent
invertible matrix $ P_{n+1,n-1}
(n \geq 1)$ over  $  {\cal R }_{n+1,n-1} ={ \cal R }\{ \,  e_1, \ \cdots, \ e_{n+1}, \  \varepsilon_1, \ \cdots, \ \varepsilon_{n-1} \, \}$ such that 
$$
 P_{n+1,n-1}( aI_{2^{n}}) P_{n+1,n-1}^{-1} \equiv \phi_{n+1,n-1}(a) \in
 {\cal R }^{2^n \times 2^n}
$$  
holds for all $ a \in {\cal R}_{n+1,n-1}$.Now let
$a \in {\cal R}_{n+2,n}={ \cal R }\{ \,  e_1, \ \cdots, \ e_{n+2}, \
\varepsilon_1, \ \cdots, \ \varepsilon_{n} \, \}. $ Then $ a $ can factor as
\begin{eqnarray*}
 a & =&  a_0 + a_1(e_{[n+1]} \varepsilon_{[n]}) + a_2(e_{[n+2]} \varepsilon_{[n-1]}) + a_3\mu_{n+2,n} \\
   & = & a_0 + (e_{[n+1]} \varepsilon_{[n]})a_1 + (e_{[n+2]} \varepsilon_{[n-1]})a_2 + \mu_{n+2,n}a_3,
\end{eqnarray*}
where 
$$
a_0, \ a_1, \ a_2, \  a_3 \in {\cal R }_{n+1,n-1} ={ \cal R }\{ \,  e_1, \ \cdots, \ e_{n+1}, \  \varepsilon_1, \ \cdots, \ \varepsilon_{n-1} \, \},
$$
$$
(e_{[n+1]} \varepsilon_{[n]})^2 = 1,  \qquad (e_{[n+2]} \varepsilon_{[n-1]})^2 =-1, 
$$ 
$$ 
\mu_{n+2,n} = (e_{[n+1]} \varepsilon_{[n]})(e_{[n+2]} \varepsilon_{[n-1]})
= -(e_{[n+2]} \varepsilon_{[n-1]})(e_{[n+1]} \varepsilon_{[n]})
= (-1)^{n+2}e_{n+2}\varepsilon_n.
$$ 
In that case$,$ $ aI_{2^{n+1}}$ satisfies the following universal similarity factorization  equality 
\begin{eqnarray*}
 \lefteqn{ P_{n+2,n}( aI_{2^{n+1}}) P_{n+2,n}^{-1} } \\
 & = & \left[ \begin{array}{cc}  \phi_{n+1,n-1}(a_0) + \phi_{n+1,n-1}(a_1)  &  -[ \, \phi_{n+1,n-1}(a_2) + \phi_{n+1,n-1}(a_3) \, ]  \\ \phi_{n+1,n-1}(a_2) - \phi_{n+1,n-1}(a_3)   & \phi_{n+1,n-1}(a_0) - \phi_{n+1,n-1}(a_1) \end{array} \right] \\
 & \equiv & \phi_{n+2,n}(a) \in {\cal R }^{2^{n+1} \times 2^{n+1}}, 
\end{eqnarray*}
where
$$
P_{n+2,n}  = \frac{1}{2} \left[ \begin{array}{cc} ( 1 + e_{[n+1]}\varepsilon_{[n]} )P_{n+1,n-1}
 &  ( e_{[n+2]}\varepsilon_{[n-1]}   - \mu_{n+2,n} )P_{n+1,n-1} \\ - ( e_{[n+2]}\varepsilon_{[n-1]} + 
\mu_{n+2,n} ) P_{n+1,n-1}  & ( 1 - e_{[n+1]}\varepsilon_{[n]} )P_{n+1,n-1} \end{array} \right], 
$$ 
$$
P_{n+2,n}^{-1}  = \frac{1}{2} \left[ \begin{array}{cc} P_{n+1,n-1}^{-1}( 1 + e_{[n]}\varepsilon_{[n]} )
 &  P_{n+1,n-1}^{-1} ( e_{[n-1]}\varepsilon_{[n-1]}   - \mu_{n+2,n} ) \\ -  P_{n+1,n-1}^{-1}( e_{[n+2]}\varepsilon_{[n-1]} + \mu_{n+2,n} )  &  P_{n+1,n-1}^{-1}( 1 - e_{[n+1]}\varepsilon_{[n]} ) \end{array} \right]. 
$$ }

The proof of this result is much analogous to that of Theorem 3.1.1. so we omit it here, and the proofs of 
next several results are also omitted . \\

\noindent {\bf 3.4. The Cases for $ {\cal R}_{n+3,n}$ } \\

For $ {\cal R}_{n+3,n}$ with $ n = 0, \ 1,$ we have the corresponding universal similarity factorization  equalities 
in Theorems 2.3.1 and 2.5.2. In general, we have the following. \\

\noindent {\bf Theorem 3.4.1.} \ {\em Let  $ a \in {\cal R}_{n+3,n}
={ \cal R }\{ \,  e_1, \ \cdots, \ e_{n+3}, \  \varepsilon_1, \ \cdots, \
\varepsilon_{n} \, \}.$ Then $ a $ can factor as 
$$
 a = a_0 + a_1e =  a_0 + ea_1,
$$
where
$$
a_0, \ a_1 \in {\cal R}_{n+2,n} = {\cal R}\{ \, e_1, \ \cdots,\  e_{n+2}, \ \varepsilon_1, \  \cdots, \ \varepsilon_n \, \},
$$
$$
e = e_{[n+3]} \varepsilon_{[n]}, \qquad e^2 = -1. 
$$ 
In that case$,$ $ aI_{2^{n+1}}$  satisfies the following universal similarity factorization  equality 
$$
 P_{n+3,n} ( aI_{2^{n+1}} ) P_{n+3,n}^{-1} = \phi_{n+2,n}(a_0) + \phi_{n+2,n}(a_1) \equiv \phi_{n+3,n}(a) \in \ {\cal C }^{ 2^{n+1} \times 2^{n+1}},  
$$ 
where  $ \phi_{n+2,n}(a_t)( t = 0, \ 1 )$ is the matrix representation of $ a_t$ in $ {\cal R}^{2^n \times 2^n }$ defined in Theorem 3.3.1 and $ P_{n+3,n} =  P_{n+2,n},$ the matrix given in Theorem 3.3.1. }\\

\noindent {\bf 3.5. The Cases for $ {\cal R}_{n+4,n}$ } \\

For $ {\cal R}_{n+4,n}$ with $ n = 0, \ 1,$ we have the corresponding universal similarity factorization  equalities 
in Theorems 2.4.1 and 2.6.2. In general, we have the following. \\

\noindent {\bf  Theorem  3.5.1.}  \ {\em Suppose that there is an independent
 invertible matrix $ P_{n+3,n-1}(n \geq 1)$ over
  $ {\cal R }_{n+3,n-1} ={ \cal R }\{ \,  e_1, \ \cdots, \ e_{n+3}, \
  \varepsilon_1, \ \cdots, \ \varepsilon_{n-1} \, \}$ such that
$$
 P_{n+3,n-1}( aI_{2^{n}}) P_{n+3,n-1}^{-1} \equiv \phi_{n+3,n-1}(a) \in {\cal H }^{2^n \times 2^n}
$$  
holds for all $ a \in {\cal R}_{n+3,n-1}$. Now let
$a \in { \cal R }_{n+4,n}={ \cal R }\{ \,  e_1, \ \cdots, \ e_{n+4}, \
\varepsilon_1, \ \cdots, \ \varepsilon_{n} \, \}.$ Then $ a $ can factor as
\begin{eqnarray*}
 a & =&  a_0 + a_1(e_{[n+4]} \varepsilon_{[n-1]}) + a_2(e_{[n+3]} \varepsilon_{[n]}) + a_3\mu_{n+4,n} \\
   & = & a_0 + (e_{[n+4]} \varepsilon_{[n-1]})a_1 + (e_{[n+3]} \varepsilon_{[n]})a_2 + \mu_{n+4,n}a_3,
\end{eqnarray*}
where 
$$
a_0, \ a_1, \ a_2, \  a_3 \in {\cal R }_{n+3,n-1} ={ \cal R }\{ \,  e_1, \ \cdots, \ e_{n+3}, \  \varepsilon_1, \ \cdots, \ \varepsilon_{n-1} \, \},
$$
$$
(e_{[n+4]} \varepsilon_{[n-1]})^2 = 1,  \qquad (e_{[n+3]} \varepsilon_{[n]})^2 =-1, 
$$ 
$$ 
\mu_{n+4,n} = (e_{[n+4]} \varepsilon_{[n-1]})(e_{[n+3]} \varepsilon_{[n]}) = -(e_{[n+3]} \varepsilon_{[n]})(e_{[n+4]} \varepsilon_{[n-1]}) = (-1)^{n+3}e_{n+4}\varepsilon_n. 
$$ 
In that case$,$ $ aI_{2^{n+1}}$ satisfies the following universal similarity factorization  equality 
\begin{eqnarray*}
 \lefteqn{P_{n+2,n}( aI_{2^{n+1}}) P_{n+2,n}^{-1} } \\
 & = & \left[ \begin{array}{cc}  \phi_{n+3,n-1}(a_0) + \phi_{n+3,n-1}(a_1)  &  -[ \, \phi_{n+3,n-1}(a_2) + \phi_{n+3,n-1}(a_3) \, ]  \\ \phi_{n+3,n-1}(a_2) - \phi_{n+3,n-1}(a_3)   & \phi_{n+3,n-1}(a_0) - \phi_{n+3,n-1}(a_1) \end{array} \right] \\
 & \equiv & \phi_{n+4,n}(a) \in {\cal H }^{2^{n+1} \times 2^{n+1}}, 
\end{eqnarray*}
where
$$
P_{n+4,n}  = \frac{1}{2} \left[ \begin{array}{cc} ( 1 + e_{[n+4]}\varepsilon_{[n-1]} )P_{n+3,n-1}
 &  ( e_{[n+3]}\varepsilon_{[n]}  - \mu_{n+4,n} )P_{n+3,n-1} \\ - ( e_{[n+3]}\varepsilon_{[n]} + 
\mu_{n+4,n} ) P_{n+3,n-1}  & ( 1 - e_{[n+4]}\varepsilon_{[n-1]} )P_{n+3,n-1} \end{array} \right], 
$$ 
$$
P_{n+4,n}^{-1}  = \frac{1}{2} \left[ \begin{array}{cc} P_{n+3,n-1}^{-1}( 1 + e_{[n+4]}\varepsilon_{[n-1]} )
 &  P_{n+3,n-1}^{-1} ( e_{[n+3]}\varepsilon_{[n]}   - \mu_{n+4,n} ) \\ -  P_{n+3,n-1}^{-1}( e_{[n+4]}\varepsilon_{[n]} + \mu_{n+4,n} )  &  P_{n+3,n-1}^{-1}( 1 - e_{[n+4]}\varepsilon_{[n-1]} ) \end{array} \right]. 
$$ }

\noindent {\bf 3.6. The Cases for $ {\cal R}_{n+5,n}$ } \\

For $ {\cal R}_{n+5,n}$ with $ n = 0, $ we have the corresponding universal similarity  factorization equality
in Theorem 2.5.1. In general, we have the following. \\

\noindent {\bf Theorem  3.6.1.} \ {\em Let  $ a \in { \cal R }_{n+5,n}
={ \cal R }\{ \,  e_1, \ \cdots, \ e_{n+5}, \  \varepsilon_1, \ \cdots, \
\varepsilon_{n} \, \}. $  Then $ a$ can factor as
$$
 a = a_0 + a_1e =  a_0 + ea_1,
$$
where
$$
a_0, \ a_1 \in {\cal R}_{n+4,n} = {\cal R}\{ \,e_1, \ \cdots,\  e_{n+4}, \ \varepsilon_1, \  \cdots, \ \varepsilon_n \, \},
$$
$$
e = e_{[n+5]} \varepsilon_{[n]}, \qquad e^2 = ( e_{[n+5]} \varepsilon_{[n]} )^2 = 1. 
$$ 
Moreover define $ \overline{a } = a_0 - a_1e $. In that case$,$ $ D_a = {\rm diag}( \, aI_{2^n}, \ 
\overline{a}I_{2^n} )$ satisfies the following universal similarity factorization  equality 
\begin{eqnarray*}
 P_{n+5,n}D_a P_{n+5,n}^{-1} & = & \left[ \begin{array}{cc} \phi_{n+4,n}(a_0) + \phi_{n+4,n}(a_1)  & O  \\
O & \phi_{n+4,n}(a_0)  - \phi_{n+4,n}(a_1)  \end{array} \right] \\
& \equiv &  \phi_{n+5,n}(a) \in \, ^2{\cal H }^{ 2^{n+1} \times 2^{n+1}},  
\end{eqnarray*} 
where  $ \phi_{n+4,n}(a_t)( t = 0, \ 1 )$ is the matrix representation of $ a_t$ in $ {\cal H}^{2^n \times 2^n }$ defined in Theorem 3.5.1 and 
$$
P_{n+5,n} = \frac{1}{2} \left[ \begin{array}{cc} (1 + e)P_{n+4,n} &
  - (1  - e)P_{n+4,n} \\ (1 - e)P_{n+4,n} & ( 1 + e)P_{n+4,n} \end{array} \right], 
$$
$$
P_{n+5,n}^{-1} = \frac{1}{2} \left[ \begin{array}{rc} P_{n+4,n}^{-1} (1 + e)
 & P_{n+4,n}^{-1}(1 - e) \\  - P_{n+4,n}^{-1}(1 - e) &  P_{n+4,n}^{-1}( 1 + e) \end{array} \right],
 $$
$ P_{n+4,n} $ and $ P_{n+4,n}^{-1} $ are the matrices given in Theorem 3.5.1. }\\

\noindent {\bf 3.7. The Cases for $ {\cal R}_{n+6,n}$ } \\

For $ {\cal R}_{n+6,n}$ with $ n = 0,$ we have the corresponding universal similarity factorization  equality 
in Theorem 2.6.1. In general, we have the following. \\

\noindent {\bf  Theorem  3.7.1.} \ {\em Suppose that there is an independent invertible matrix $ P_{n+5,n-1}
(n \geq 1)$ over  $  {\cal R }_{n+5,n-1} ={ \cal R }\{ \,  e_1, \ \cdots, \ e_{n+5}, \  \varepsilon_1, \ \cdots, \ \varepsilon_{n-1} \, \}$ such that 
$$
 P_{n+5,n-1}( aI_{2^{n+1}}) P_{n+5,n-1}^{-1} \equiv \phi_{n+5,n-1}(a) \in {\cal H }^{2^{n+1} \times 2^{n+1}}
$$  
holds for all $ a \in {\cal R}_{n+5,n-1}$. Now
let $ a \in {\cal R}_{n+6,n}={\cal R}\{ \,  e_1, \ \cdots, \ e_{n+6}, \
\varepsilon_1, \ \cdots, \ \varepsilon_{n} \, \}.$ Then $ a$ can factor as
\begin{eqnarray*}
 a & =&  a_0 + a_1(e_{[n+5]} \varepsilon_{[n]}) + a_2(e_{[n+6]} \varepsilon_{[n-1]}) + a_3\mu_{n+6,n} \\
   & = & a_0 + (e_{[n+5]} \varepsilon_{[n]})a_1 + (e_{[n+6]} \varepsilon_{[n-1]})a_2 + \mu_{n+6,n}a_3,
\end{eqnarray*}
where 
$$
a_0, \ a_1, \ a_2, \  a_3 \in {\cal R }_{n+5,n-1} ={ \cal R }\{ \,  e_1, \ \cdots, \ e_{n+5}, \  \varepsilon_1, \ \cdots, \ \varepsilon_{n-1} \, \},
$$
$$
(e_{[n+5]} \varepsilon_{[n]})^2 = 1,  \qquad (e_{[n+6]} \varepsilon_{[n-1]})^2 =-1, 
$$ 
$$ 
\mu_{n+6,n} = (e_{[n+5]} \varepsilon_{[n]})(e_{[n+6]} \varepsilon_{[n-1]}) = -(e_{[n+6]} \varepsilon_{[n-1]})(e_{[n+5]} \varepsilon_{[n]}) = (-1)^{n+5}e_{n+6}\varepsilon_n. 
$$ 
In that case$,$ $ aI_{2^{n+2}}$ satisfies the following universal similarity  factorization equality
\begin{eqnarray*}
 \lefteqn{ P_{n+6,n}( aI_{2^{n+2}}) P_{n+6,n}^{-1} } \\
 & = & \left[ \begin{array}{cc}  \phi_{n+5,n-1}(a_0) + \phi_{n+5,n-1}(a_1)  &  -[ \, \phi_{n+5,n-1}(a_2) + \phi_{n+5,n-1}(a_3) \, ]  \\ \phi_{n+5,n-1}(a_2) - \phi_{n+5,n-1}(a_3)   & \phi_{n+5,n-1}(a_0) - \phi_{n+5,n-1}(a_1) \end{array} \right] \\
& \equiv & \phi_{n+6,n}(a) \in {\cal H }^{2^{n+2} \times 2^{n+2}}, 
\end{eqnarray*}
where
$$
P_{n+6,n}  = \frac{1}{2} \left[ \begin{array}{cc} ( 1 + e_{[n+5]}\varepsilon_{[n]} )P_{n+5,n-1}
 &  ( e_{[n+6]}\varepsilon_{[n-1]}  - \mu_{n+6,n} )P_{n+5,n-1} \\ - ( e_{[n+6]}\varepsilon_{[n-1]} + 
\mu_{n+6,n} ) P_{n+5,n-1}  & ( 1 - e_{[n+5]}\varepsilon_{[n]} )P_{n+5,n-1} \end{array} \right], 
$$ 
$$
P_{n+6,n}^{-1}  = \frac{1}{2} \left[ \begin{array}{cc} P_{n+5,n-1}^{-1}( 1 + e_{[n+5]}\varepsilon_{[n]} )
 &  P_{n+5,n-1}^{-1} ( e_{[n+6]}\varepsilon_{[n-1]}   - \mu_{n+6,n} ) \\ -  P_{n+5,n-1}^{-1}( e_{[n+6]}\varepsilon_{[n-1]} + \mu_{n+6,n} )  &  P_{n+5,n-1}^{-1}( 1 - e_{[n+5]}\varepsilon_{[n]} ) \end{array} \right]. 
$$ }

In the same manner it is not difficult to give the induction formulas for the universal similarity factorization  equalities 
over $ {\cal R}_{n, n+q}$ with $ 1 \leq q \leq 6,$  we leave them to the reader. As to
 $ {\cal R}_{n+7,n}$ and $ {\cal R}_{n, n+7}$ with $ n = 0$, the corresponding universal similarity 
factorization  equalities have been given in Subsection 2.7, the general results corresponding to nonzero 
$ n $ will be included in the next two sections. \\

\noindent {\bf 4.  UNIVERSAL SIMILARITY EQUALITIES OVER $ {\cal R}_{p+8,0} $ AND $ {\cal R}_{0,q+8} $ } \\

According to the two formulas in (1.9), we know that
$$
 {\cal R}_{p+8,0}  \ \simeq \  {\cal R}_{p,0}^{16 \times 16}, \qquad {\cal R}_{0,q+8} \  \simeq
  \ {\cal R}_{0,q}^{16 \times 16} 
$$ 
hold  for all  finite $ p $ and $ q $.  Now applying the theorems in Subsection 2.8, we have the following
 two general results on the  universal similarity  factorization equalities over $ {\cal R}_{p+8,0} $ and $ {\cal R}_{0,q+8}.$ \\

\noindent {\bf Theorem 4.1.} \ {\em Let
$$
a \in {\cal R}_{p+8,0}
= {\cal R}\{ e_1, \ \cdots, \ e_8, \ \alpha_1, \
 \cdots, \  \alpha_p \  | \ e_i^2 = \alpha_j^2 =1 , \ i = 1, \ \cdots, \ 8,
 \  j = 1, \ \cdots, \ p \}.
$$
Then $ a $ can factor as
$$
a = \sum_A a_A (e_{[8]}\alpha )_A \in {\cal R}_{8,0}\{ \, e_{[8]}\alpha_1, \
 \cdots, \ e_{[8]}\alpha_p \  | \ ( e_{[8]}\alpha_j)^2  =1 , \ j = 1, \
 \cdots, \ p \, \},
$$ 
where $ A = ( \, j_1, \ j_2, \ \cdots,  \ j_k \,  ) $ with
$ 1 \leq j_1 < j_2 < \cdots < j_k \leq  p$ ranging all naturally ordered
subsets of $ \{ 1, \ 2, \ \cdots, \ p \};$  $a_A$ has the form
$$
a_A \in {\cal R}_{8,0} = {\cal R}\{ \, e_1, \ \cdots, \ e_8 \, \};
$$
$ (e_{[8]}\alpha )_A $
is  defined to be
$$
(e_{[8]}\alpha )_A = (e_{[8]}\alpha )_{(j_1, j_2, \cdots, j_k)} 
 \equiv (e_{[8]}\alpha_{j_1}) (e_{[8]}\alpha_{j_2}) \cdots (e_{[8]}\alpha_{j_p}), \qquad
(e_{[8]}\alpha )_{A = \emptyset} \equiv e_{[8]};  
$$ 
both of which satisfy
$$
a_A (e_{[8]}\alpha )_A  = (e_{[8]}\alpha )_Aa_A.
$$
In that case$,$ $ aI_{16}$ satisfies the following universal similarity
factorization equality
$$
 P_{8,0}(aI_{16})P_{8,0}^{-1} = \sum_A  \phi_{8,0}(a_A)(e_{[8]}\alpha )_A \equiv \phi_{p+8,0}(a), 
  \eqno (4.1)
$$ 
where $  \phi_{8,0}(a_A)$  is the matrix representation of $ a_A$ in $ {\cal R}^{16 \times 16}$ defined
 in Theorem 2.8.1, $ P_{8,0} $ is the independent invertible matrix mentioned  in Theorem 2.8.1,
meanwhile,
$$
\phi_{p+8,0}(a) \in { \cal R }^{16 \times 16}  \left\{ \, e_{[8]}\alpha_1, \
 \cdots, \  e_{[8]}\alpha_p \ | \ (e_{[8]}\alpha_j)^2 = 1, \ j = 1, \ \cdots,
  \ p \, \right\} = { \cal R }^{16 \times 16}_{p,0},
   \eqno (4.2)
$$
called the matrix representation of $ a \in {\cal R}_{p+8,0}$ in ${ \cal R }^{16 \times 16}_{p,0}$.} \\

If $ 1 \leq p \leq 8$ in ${\cal R}_{p+8,0}$, applying the results in Section
 2 to the $16 \times 16 $ matrix  $ \phi_{p+8,0}(a)$ in Eq.(4.1) we can
 establish a universal similarity factorization
 equality between elements of  $ {\cal R}_{p+8,0} $  and matrices
 with elements in  $ {\cal R}$, or $ {\cal C}$,
 or ${\cal H}.$  \\

Similarly we  have  the following. \\

\noindent {\bf Theorem  4.2.} \ { \em Let
$$
a \in {\cal R}_{ 0,q+8} = {\cal R}\{ \tau_1, \ \cdots, \ \tau_8, \
\varepsilon_1, \ \cdots, \ \varepsilon_q \  | \
\tau_i^2 = \varepsilon_j^2 =-1 , \ i = 1, \ \cdots, \ 8, \  j = 1, \ \cdots,
 \ q \}.
$$
Then $ a $ can factor as   
$$
a = \sum_A a_A ( \tau_{[8]}\varepsilon)_A  \in  {\cal R}_{0,8}\{ \tau_{[8]} \varepsilon_1, \ \cdots, \ \tau_{[8]} \varepsilon_q \  | \ ( \tau_{[8]}\varepsilon_j)^2  = -1 , \ j = 1, \ \cdots, \ q \},
$$ 
where $ A = ( \, j_1, \ j_2, \ \cdots,  \ j_k \,  ) $ with $ 1 \leq j_1 < j_2 < \cdots < j_k \leq q$ ranges all naturally ordered subsets of $ \{ 1, \ 2, \ \cdots, \ q \};$  $ (\tau_{[8]}\varepsilon)_A $ and $ a_A$ are 
$$ 
(\tau_{[8]}\varepsilon)_A  = (\tau_{[8]}\varepsilon )_{(j_1, j_2, \cdots, j_k)} 
 \equiv (\tau_{[8]}\varepsilon_{j_1}) (\tau_{[8]}\varepsilon_{j_2}) \cdots (\tau_{[8]}\varepsilon_{j_k}), 
\qquad (\tau_{[8]}\varepsilon)_{A = \emptyset} \equiv \tau_{[8]},     
$$ 
$$
a_A \in {\cal R}_{0,8} = {\cal R}\{ \, \tau_1, \ \cdots, \ \tau_8, \ | \
\tau_i^2 = -1, \ i = 1, \ \cdots,
\ 8 \, \}, 
$$
and both of them satisfy
$$
a_A (\tau_{[8]}\varepsilon )_A  = (\tau_{[8]}\varepsilon )_A a_A.
$$
In that case$,$ $ aI_{16}$  satisfies the following universal similarity factorization  equality 
$$
 P_{0,8}(aI_{16})P_{0,8}^{-1} = \sum_A  \phi_{0,8}(a_A)(\tau_{[8]} \varepsilon)_A \equiv \phi_{0,q+8}(a),  \eqno (4.3)
$$ 
where $  \phi_{0,8}(a_A)$  is the matrix representation of $ a_A$ in $ {\cal R}^{16 \times 16}$ defined
 in Theorem 2.8.2, $ P_{0,8} $ is the independent invertible matrix mentioned in Theorem 2.8.2,
meanwhile,
$$
\phi_{0,q+8}(a) \in { \cal R }^{16 \times 16} \left\{ \,
\tau_{[8]}\varepsilon_1, \  \cdots, \  \tau_{[8]}\varepsilon_q \ |
\ (\tau_{[8]}\varepsilon_j)^2 = -1, \ j = 1,  \ \cdots, \ q \, \right\}
 = { \cal R }^{16 \times 16}_{0,q},   \eqno (4.4)
$$
called the matrix representation of $ a \in {\cal R}_{0,q+8}$ in ${ \cal R }^{16 \times 16}_{0,q}$.} \\

If $ 1 \leq q \leq 8$ in $ {\cal R}_{0,q+8}$, then  by applying the results in Section 2 to the  
$16 \times 16 $ matrix  $ \phi_{0,q+8}(a)$ in Eq.(4.3) we can establish a universal similarity
 factorization  equality  between elements of  $ {\cal R}_{0,q+8} $  and matrices over $ {\cal R}$ or $ {\cal C}$
 or ${\cal H}.$  \\

\noindent {\bf 5.  UNIVERSAL SIMILARITY EQUALITIES OVER $ {\cal R}_{p+8,q} $ AND $ {\cal R}_{p,q+8} $ } \\

For the two algebraic isomorphisms in Eq.(1.8), we have the following two general results.\\

\noindent {\bf Theorem  5.1.} \ {\em Let $ a \in {\cal R}_{p+8,q}= {\cal R}\{ \, e_1, \ \cdots, \ e_8, \ \alpha_1, \
 \cdots, \  \ \alpha_p, \ \varepsilon_1, \ \cdots, \ \varepsilon_q \   | \ e_i^2 = \alpha_j^2 =
 - \varepsilon_k^2 =1, \ i = 1, \ \cdots, \ 8, \  j = 1, \ \cdots, \ p, \ k =
  1, \ \cdots, \ q \, \}.$ Then $ a $ can factor as   
$$
a = \sum_A a_A (e_{[8]}\alpha\varepsilon)_A  \in  {\cal R}_{8,0}\{ \ e_{[8]}\alpha_1, \ \cdots, \ e_{[8]}\alpha_p, \ e_{[8]} \varepsilon_1, \ \cdots, \ e_{[8]}\varepsilon_q \ \},
$$ 
where 
$$ 
( e_{[8]}\alpha_j)^2  = 1 , \qquad ( e_{[8]}\varepsilon_k )^2 =-1, \qquad  j = 1, \ \cdots, \ \ p, \qquad k = 1, \ \cdots, \ q;
$$
$ A =( \, A_1, \ A_2 \,) $ is the combination of  the two ordered multiindices $ A_1$ and $ A_2$  
$$ 
 A_1 = ( \, j_1, \ j_2, \ \cdots,  \ j_s \, ), \ \ \ \ A_2 = ( \, k_1, \ k_2, \ \cdots,  \ k_t \, ),  
$$ 
with $ 1 \leq j_1 < j_2 < \cdots < j_s \leq  p$ and $ 1 \leq k_1 < k_2 < \cdots < k_t \leq  q$ ranging all 
natually ordered subsets of $ \{ 1, \ 2, \ \cdots, \ p \}$ and $ \{ 1, \ 2, \ \cdots, \ q \},$ 
respectively$;$
$$ 
(e_{[8]}\alpha \varepsilon)_A \equiv (e_{[8]}\alpha )_{A_1}( e_{[8]}\varepsilon)_{A_2}=
 (e_{[8]}\alpha_{j_1}) (e_{[8]}\alpha_{j_2}) \cdots (e_{[8]}\alpha_{j_s}) (e_{[8]}\varepsilon_{k_1}) (e_{[8]}\varepsilon_{k_2}) \cdots (e_{[8]}\varepsilon_{k_t}), 
$$
$$
(e_{[8]}\alpha \varepsilon)_{A = \emptyset} \equiv e_{[8]},  
$$
$$ 
a_A \in {\cal R}_{8,0} = {\cal R}\{ \, e_1, \ \cdots, \ e_8 \ | \ e_i^2 = 1,
 \ i = 1, \ \cdots, \ 8 \, \},
$$
with
$$
a_A (e_{[8]}\alpha\varepsilon )_A  = (e_{[8]}\alpha\varepsilon )_A a_A
$$
always holding. In that case$,$ $ aI_{16}$  satisfies the following  universal similarity  factorization equality 
$$
 P_{8,0}(aI_{16})P_{8,0}^{-1} = \sum_A  \phi_{8,0}(a_A)(e_{[8]}\alpha \varepsilon)_A \equiv \phi_{p+8,q}(a),   \eqno (5.1)
$$ 
where $  \phi_{8,0}(a_A)$  is the matrix representation of $ a_A$ in $ {\cal R}^{16 \times 16}$ defined
 in Theorem 2.8.1, $ P_{8,0} $ is the independent invertible matrix mentioned in Theorem 2.8.1,
meanwhile,
$$
\phi_{p+8,q}(a) \in { \cal R }^{16 \times 16} \{ \, e_{[8]}\alpha_1, \  \cdots, \  e_{[8]}\alpha_p,
 \  e_{[8]}\varepsilon_1, \  \cdots, \  e_{[8]}\varepsilon_q \} 
= { \cal R }^{16 \times 16}_{p,q}. \eqno (5.2)
$$ } 

If $ 1 \leq p \leq 8$ and $ 1 \leq q \leq 8$ a in $ {\cal R}_{p+8,q}$, then  by applying the results in Section 2 to the  $16 \times 16 $ matrix  $ \phi_{p+8,q}(a)$ in Eq.(5.1) we can  establish a universal similarity factorization  equality between elements of  $ {\cal R}_{p+8,q} $  and matrices over $ {\cal R}$ or $ {\cal C}$ or ${\cal H}.$  \\

The result for $ {\cal R}_{p,q+8} $ is  much similar to that of  Theorem 5.1, so  we omit it here. \\

\end{document}